\documentclass{aa}

\usepackage{graphicx}
\usepackage{caption}
\usepackage{txfonts}
\usepackage[normalem]{ulem}
\begin{document}

   \title{Asymmetric emission of the [OIII]$\lambda$5007 profile in narrow-line Seyfert 1 galaxies \thanks{The reduced spectra as FITS files are only available in electronic form at the CDS via anonymous ftp to cdsarc.u-strasbg.fr (130.79.128.5)
or via http://cdsweb.u-strasbg.fr/cgi-bin/qcat?J/A+A/}
   }

   \author{E. O. Schmidt
          \inst{1}\fnmsep\thanks{Visiting Astronomer, Complejo 
Astron\'omico El Leoncito operated under agreement between the Consejo Nacional de Investigaciones Cient\'ificas y T\'ecnicas de la Rep\'ublica Argentina and the National Universities of La Plata, C\'ordoba and San Juan.}
          \and
          G. A. Oio\inst{1}
          \and
           D. Ferreiro\inst{1,2}
           \and
         L. Vega\inst{1,2}
          \and
                W. Weidmann \inst{2,3}
        }

   \institute{Instituto de Astronomía Te\'orica y Experimental (IATE), Universidad Nacional de C\'ordoba, CONICET, Observatorio Astron\'omico de C\'ordoba, Laprida 854, X5000BGR, C\'ordoba, Argentina. \\
              \email{eduschmidt@oac.unc.edu.ar}
         \and
            Observatorio Astron\'omico de C\'ordoba, Laprida 854, X5000BGR, C\'ordoba, Argentina.
           \and 
           Consejo de Investigaciones Cient\'{i}ficas y T\'ecnicas (CONICET) de la Rep\'ublica Argentina.
           }


 
  \abstract
   {Many active galactic nuclei (AGN) and particularly narrow-line Seyfert 1 (NLS1) galaxies, usually exhibit blueshifts and blue wings in several emission lines, which are mainly associated with outflows and strong winds. Although there is no clear evidence of the physical origin of the [OIII] blue wings, it has been suggested that they could be emitted from the inner parts of the narrow-line region (NLR).}
   {In order to study the radial velocity difference between the narrow component of H$\beta$ and the core component of [OIII]$\lambda$5007 and the asymmetric emission of this forbidden line, we investigate a sample of NLS1 galaxies . One of the aims of this paper is to analyze the blue wings of the [OIII]$\lambda$5007 profiles and their relation with the central engine.}
   {We have obtained and studied medium-resolution spectra (190 km s$^{-1}$ FWHM at H$\beta$) of a sample of 28 NLS1 galaxies in the optical range 4300 - 5200\AA. We performed Gaussian decomposition to the H$\beta$ and [OIII]$\lambda\lambda$4959,5007 emission profiles in order to study the distinct components of these lines.}
   {A new blue outlier galaxy is found, in which the center of the core component of [OIII] is blueshifted by 405 km s$^{-1}$ relative to the center of the narrow component of H$\beta$ line. We confirmed a previously known correlation between the blueshift and the full width half maximum (FWHM) of the core component of [OIII]$\lambda$5007 line. We also corroborated the correlation between the latter and the velocity of the centroid of the blue wing. On the other hand, by studying the radial velocity difference between the blue end of the asymmetric emission and the centroid of the core component of [OIII], we found a correlation between it and the central black hole mass and, therefore, with the luminosity of the broad component of H$\beta$. Finally, we found a moderate correlation between the luminosity of the [OIII] blue wing and the black hole mass.}
   {These results suggest that the asymmetric emission of the [OIII] lines is related to the central engine, not only through the black hole but also in the intensity of the AGN, which is in agreement with previous results.}

  \keywords{Galaxies: active --
                nuclei --
                Seyfert --
                kinematics and dynamics
               }

  \authorrunning{Schmidt et al.}\space\maketitle
%
%

\section{Introduction}

        Narrow-line Seyfert 1 (NLS1) galaxies are a subclass of AGN defined according to the width of the broad component of their optical Balmer emission lines and their relative weakness of the [OIII]$\lambda$5007 emission with full width at half maximum (FWHM) $<$ 2000 km s$^{-1}$ and [OIII]$\lambda$5007/H$\beta$ $\leq$3 \citep{1985ApJ...297..166O,1989ApJ...342..224G}. Classifications based on optical spectroscopy suggest that NLS1 represent about 15\% of the whole population of Seyfert 1 galaxies \citep{2002AJ....124.3042W}.   

        Related to the nuclear kinematics, the presence of blue wings and blueshifts in several high and medium degree of ionization emission lines have been reported in different AGNs and were associated with winds and outflows \cite[e.g.,][]{1987PASP...99..809B,2003ApJS..145..199M}. These outflows can provide energy and mass to the interstellar medium of the host galaxy \citep{1996ApJS..105...75C,2001ApJ...554..261C,2007A&A...463..513M}. 
    
    There is observational evidence of these blueshifts in all wavelengths from the X-ray band \citep[e.g.,][]{2005ApJ...625...95C,2007ApJ...659.1022K} to UV \citep[e.g.,][]{2007ApJ...659..250C,2007ApJ...666..757S}, optical \citep[e.g.,][]{2005AJ....130..945D}, IR \citep[e.g.,][]{RA2006}, and radio \citep[e.g.,][]{2006AJ....132..546G}.
        
        \cite{2002ApJ...576L...9Z} studied a sample of 216 type 1 AGNs and found 7 of these sources with [OIII]$\lambda$5007 blueshifts $< -$250 km s$^{-1}$ (called blue outliers). Likewise, objects with similar characteristics were analyzed by several authors \citep[e.g.,][]{2001A&A...369..450G,2002xsac.conf..287G, 2003MNRAS.345.1133M, 2005ApJ...618..601A}. Recently, \cite{2016MNRAS.459.3144Z} detected quasars at z$\sim$2.5 with blueshifts of $\sim -$1500 km s$^{-1}$.
        
Considering high ionization emission lines such as [Fe VII], [Fe X], and [Fe XI], \cite{RA2006} analyzed blueshifts in Seyfert 1 galaxies. They found that the size of the emitting region varies with the ionization potential. In this scenario, the higher the ionization potential, the more compact the emitting region becomes. This suggests that nuclear photoionization is the principal excitation mechanism. However, these authors suggest that shocks generated by the outflow could provide an additional amount of energy for line formation.

In a study of NLS1 galaxies, a blueshift of $\sim$10 $\AA{}$ ($\sim$600 km s$^{-1}$) was measured in the [OIII]$\lambda$5007 emission profile of the galaxy Zw1 \citep{1987PASP...99..809B}. In addition, \cite{2005ApJ...618..601A} found two NLS1 galaxies with [OIII] blueshifts of $\sim$1000 km s$^{-1}$ relative to H$\beta$. These objects also present line widths of 1000$-$2000 s$^{-1}$. These authors suggest that outflows take place in their nuclei and interact with the ambient gas. \cite{2005MNRAS.364..187B} studied a sample of 150 Sloan Digital Sky Survey NLS1 galaxies and found 7 blue outlier objects and a correlation between the [OIII] blueshift and the FWHM of the core component of [OIII] line. One of the aims of this paper is to study the blueshifts of [OIII]$\lambda$5007 respect to H$\beta$ in NLS1 galaxies to investigate the dynamics of the narrow-line region (NLR) and to detect possible blue outlier galaxies.

On the other hand, and related to the blue wings, NLS1 galaxies often show blue asymmetric profiles in the [OIII]$\lambda\lambda$4959,5007 lines \citep[e.g.,][]{1981ApJ...247..403H,1985ApJ...294..106V}. These asymmetric profiles are mainly composed by a core component that is emitted from the NLR, which has typical values of FWHM of a few hundred km s$^{-1}$, and a blue wing with higher FWHM ($\sim$500 km s$^{-1}-$ 1500km s$^{-1}$), which originated closer to the active nucleus \citep{2001A&A...372..730V,2005MNRAS.364..187B,2008K}. Some studies suggest that the difference of velocities between the core and asymmetric component of [OIII] lines correlates with the FWHM of these lines \citep{2016MNRAS.462.1256C,2011ApJ...739...28X}. In this paper we study the FWHM and luminosity of the blue wing emission and their possible connection with the central black hole.

    In order to study the blueshifts and blue wings and their relation with the central engine, we studied 28 NLS1 galaxies with marked blue wings in the [OIII]$\lambda$5007 emission line. Of these objects, 22 galaxies lie in the southern hemisphere. We carried out a spectroscopic study of these objects to analyze the outflows and the asymmetries of the profiles. We present the sample and the observations in  Sect. \ref{observation}, the measurement process of the emission lines are described in Sect. \ref{measurements}. In Sect. \ref{sec:oiii}, the blueshifts and blue wings of [OIII]$\lambda$5007 and their relation with the black holes are studied. In Sect. \ref{sec:discussion} we analyze the results and finally, we draw our conclusions in Sect. \ref{final}. Throughout this paper, we use the cosmological parameters $H_{0} = 70$ Km s$^{-1}$ Mpc$^{-1}$, $\Omega_{M}= 0.3$, and $\Omega_{\Lambda}= 0.7$.

    \section{The sample and observations}
    \label{observation}
    
    For this work, we selected the 20 NLS1 galaxies of the sample of \cite{Schmidt}, which exhibit marked blue wings in the [OIII]$\lambda$5007 emission line. The objects were originally selected from the Véron \& Véron catalog \citep{2010A&A...518A..10V} whose nuclear kinematics have been poorly or even not studied; these objects have redshifts z$<$0.15 and $\delta \le 10^{\circ}$. We added 8 objects from the same catalog, 6 of which have the same characteristics as the main sample and 2  have z$=$0.1547 and z$=$0.2063. The final sample consists of 28 poorly studied NLS1 galaxies (most of them from the southern hemisphere) with an evident asymmetry in the [OIII]$\lambda$5007 emission line. Table \ref{tab:sample} lists the galaxy name, right ascension, declination, radial velocity, redshift, apparent magnitude and filter, major diameter, and minor diameter. The data were taken from the Nasa Extragalactic Database (NED). The object MCG$-$04.24.017 is the only source with morphological classification, as it is classified as a SB0 galaxy \citep{1991rc3..book.....D}.

    Observations were performed in different campaigns between 2011 and 2015 using the REOSC Spectrograph at 2.15 m. telescope of the Complejo AStronómico el LEOncito (CASLEO), in Argentina. The spectrograph has attached a Tektronix 1024 $\times$ 1024 CCD with 24 $\mu$m pixels. The galaxies were observed using a 2.7 arcsec wide slit and the extractions of each spectrum were of $\sim$ 2.3 arcsec. We used a 600 line mm$^{-1}$ grating giving a resolution of 190 km \ s$^{-1}$ FWHM around H$\beta$.  
      For this paper we obtained spectra in the blue range 4300\AA \ to 5200\AA \ for the whole sample. In addition, we also obtained spectra for 6 of the additional galaxies in the red range 5800\AA \ to 6800\AA, with a resolution of 170 km \ s$^{-1}$ FWHM around H$\alpha$. Table \ref{tab:obs} lists the observational data for the sample, such as the date of the observation and the exposure time of the blue and red spectra. Figure \ref{fig:spec1} presents the spectra for the 28 NLS1 galaxies in the blue spectral region 4300\AA \ to 5200\AA. All spectra are in rest frame considering the peak of the [OIII]$\lambda$5007 profile. In this paper, we focus on this spectral range in order to study H$\beta$ and [OIII]$\lambda$5007 emission lines in detail. In Fig. \ref{fig:sper}, we show the spectra for 6 additional galaxies in the red spectral range 5800\AA \ to 6800\AA. In this way, we can measure emission lines such as H$\alpha+$[NII]$\lambda\lambda$6548,6584 and estimate the black hole masses (see Sect. \ref{totalasy}).\\
      For a mean distance of 240 Mpc (mean redshift of 0.058) the obtained spectra correspond to a central projected distance of 3 kpc. According to some previous studies, the host contribution emission in NLS1 galaxies is around 30\% \citep[e.g.,][]{2006ApJS..166..128Z}. Therefore, some of the studied emission lines in this work could be contaminated. On the other hand, the outflows seem to be constrained within radial spacial scales of few kpc from the central engine \citep{Villar-Martin}. The high excitation suggests that stars are not responsible for ionizing the outflowing gas, but rather AGN related phenomena. These studies claimed that the outflows are within the AGN ionization cones. \cite{Rice} suggested that the asymmetries are mainly due to nuclear emitting gas, rather than material distributed throughout the NLR on larger scales. Considering that the broad component of H$\beta$ is originated in the BLR and taking into account what we mentioned above, the only emission line of interest in our present work that could be contaminated, is the core component of [OIII]. Following \cite{2006ApJS..166..128Z}, this contamination could be around 30\%.

    \section{Measurements}
    \label{measurements}

    \subsection{Iron emission lines}
    \label{fe}
    NLS1 galaxies usually exhibit, on average, strong optical Fe emission lines in the vicinity of H$\beta$ and [OIII]$\lambda$5007. These objects exhibit a wider range of iron emission compared with broad line AGNs \citep[e.g.,][]{2014Natur.513..210S}. According to the lower term of their transitions, these Fe lines are of the S group. This group consists of several lines, which correspond to multiplets 41, 42, 43. The lines of the multiplet 42 overlap with H$\beta$ and [OIII]$\lambda\lambda$4959,5007 profiles, which are the lines of  interest in this work. Because of this, we subtracted the iron emission using the online software\footnote{\url{http://servo.aob.rs/FeII_AGN/}} developed by \cite{2010ApJS..189...15K} and \cite{2012ApJS..202...10S}. This software provides a best-fit model that reproduces the iron multiplets in the H$\beta$ region for each object in function of the gas temperature, Doppler broadening, and shift of the FeII lines. As a first approximation we used the initial parameters provided in the web page. These fits were visually inspected and the initial parameters were modified until acceptable fits were obtained. 
    Some galaxies such as 1RXS J040443.5$-$295316, 2MASX J01413249$-$1528016, 2MASXJ21565663$-$1139314, CTSH34.06, CTSM02.47, and Zw049.106 do not show Fe emission and the fit was not necessary. The fe multiplet fitting was not satisfactory in RX J0024.7$+$0820, SDSSJ161227.83+010159.8, and SDSS J225452.22$+$004631.4, probably due to the low spectral S/N, and we decided not to take this fitting into account. The left panels of Fig. \ref{fig:spec1} show the Fe fits on the continuum subtracted spectra of the galaxies.
    
    \subsection{Emission line measurements}
    \label{emissionlines}
    
    We are interested in the emission lines, which provide important information about the kinematics of the central regions of AGNs. Emission lines in NLS1 galaxies can be represented by a single or a combination of Gaussian functions \citep[e.g.,][]{Xu2007,Villar-Martin,Schmidt,2017ApJ...848...35S}. For this purpose, we used the LINER routine \citep{1993Pogge..and..Owen}, which is a $\chi^2$ minimization algorithm that can fit several Gaussian functions to a line profile.

        At first, we fitted the continuum in the vicinity of the lines of interest. The fit procedures for the  H$\alpha +$[NII]$\lambda$6548, 6584 profiles, and [SII]$\lambda\lambda$6716,6731 emission lines are detailed in \cite{Schmidt}. 
    
    We fitted the [OIII]$\lambda$5007 emission line with one Gaussian function for the core component ([OIII]$_{cc}$) and with another Gaussian for the asymmetric component \citep[e.g.,][]{2008K}. To fit the narrow component (NC) of H$\beta$ we had taken into account that it should have approximately the same FWHM as [OIII]$_{cc}$ \citep[e.g.,][]{Villar-Martin,2017ApJ...848...35S}. For the broad component (BC) of H$\beta$, we fitted one or two Gaussian functions \citep[e.g.,][]{2008K}, depending on the case. We followed the same criterion as in \cite{Schmidt} of minimizing the residuals obtained with the smaller number of Gaussian functions. All the residuals were visually inspected.
   
    In order to estimate the uncertainties, we measured 15 times some galaxies with different S/N (for example, S/N of around 5, 13, and 22). We assumed that the errors are given by the dispersion in the distribution of the measurements, where we adopted them at 1$\sigma$. The relative errors of the flux and FWHM of the broad component of H$\beta$ and H$\alpha$ are in the ranges 4\%$-$10 \% and 5\%$-$20\%, respectively. The error for the flux and FWHM of the core component of [OIII] are in the ranges 3\%$-$12\% and 1\%$-$ 5\%, respectively. For the asymmetric component of [OIII], the relative error is 2\%$-$10\% for the flux and 4\%$-$15\% for the FWHM. Nevertheless, these uncertainties concern only the measurement process and are a lower limit of the global errors. We did not consider, for example, the uncertainties in the observation system or calibrations. The narrow component of H$\beta$ and the core component of [OIII] are characterized by high covariance.
    
    As mentioned, strong optical iron emission lines are usually observed in NLS1 spectra, therefore, to see how much the Fe multiplet fitting procedure affects measurements of [OIII] fluxes, FWHM, and asymmetries, we considered some galaxies with an intense iron emission (for example, FAIRALL107, RBS02.19, and RHS56) and we repeated the spectral analysis several times under two different extreme conditions. We fit both the observed and FeII subtracted spectra and we obtained two different mean values for the flux and  FWHM of the narrow and the broad component of H$\beta$ and the core and asymmetric component of [OIII] line. Under this extreme condition and considering high iron emitting galaxies, we obtained that the variations in the FWHM and flux of the narrow and broad components of H$\beta$ are less than 5\%. The variation of the FWHM and flux of the core component of [OIII] line is 1\%$-$4\% and 3\%$-$12\%, respectively. As we expected, the higher variation is seen considering the asymmetric component of [OIII] line, being of 5\%$-$10\% and 10\%$-$20\%, for the FWHM and the flux, respectively.
    
\begin{figure}[!ht]
 \begin{minipage}{\linewidth}
 \begin{center}
 \includegraphics[trim = 0mm 00mm 0mm 00mm, clip, width=1\textwidth]{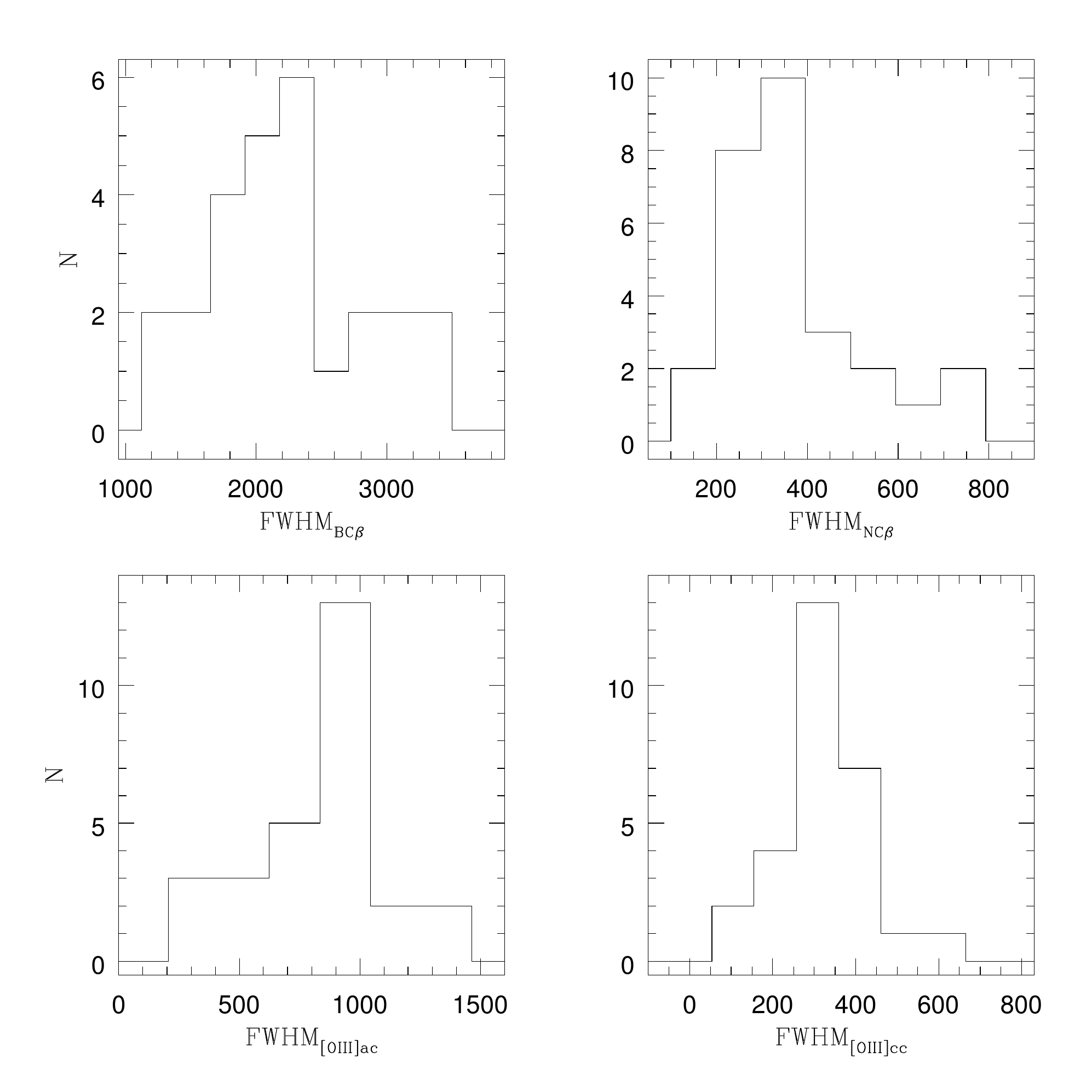}
  \end{center}
 \caption{FWHM distributions of the emission lines in km  s$^{-1}$. From left to right and top to bottom: FWHM of BC$\beta$, NC$\beta$, [OIII]$_{ac}$, and [OIII]$_{cc }$ are shown. All FWHM were corrected by the instrumental width.}
\label{fig:hist.fwhm}
 \end{minipage}
\end{figure}

     Figure \ref{fig:hist.fwhm} shows the distributions of FWHM of BC$\beta$, NC$\beta$, [OIII]$_{cc}$, and [OIII]$_{ac}$. It can be seen that most of the galaxies have FWHM of BC$\beta$ of $\sim$2000 km s$^{-1}$ (top left panel). The top right panel shows that the FWHM of NC$\beta$ is in the range of 200$-$800 km s$^{-1}$. The same range of velocities is shown by [OIII]$_{cc}$ in the bottom right panel, and finally, the FWHM of [OIII]$_{ac}$ is in the range of 300$-$1400 km s$^{-1}$. All values are in agreement with previous measurements of lines emitted at the BLR and NRL \citep[e.g.,][]{Oster,2000ApJ...538..581R,Schmidt} and in concordance with the known values of the blue wings of the [OIII]$\lambda$5007 emission lines \citep[e.g.,][]{2005MNRAS.364..187B}. All the presented FWHM were corrected by the instrumental width as FWHM$^{2}$ $=$ FWHM$_{obs}^{2}$ $-$FWHM$_{inst}^{2}$, where FWHM$_{obs}$ is the measured FWHM and FWHM$_{inst}$ is the instrumental broadening ($\sim$3\AA\ or 190 km s$^{-1}$).

    \section{ Emission profile of [OIII]$\lambda$5007 }
    \label{sec:oiii}
    
    \subsection{Blueshifts}
    \label{blueshift}
    
    In order to determine the blueshift of [OIII], we measured the wavelength difference between the centroid of the narrow component of H$\beta$ (at $\lambda =$4861.3 $\AA{}$) and the core component of [OIII] (at $\lambda =$5006.8 $\AA{}$). Hence, the blueshift of [OIII] relative to H$\beta$ can be calculated based on the laboratory wavelength difference of both emission lines (145.5 $\AA{}$). The uncertainties of the determinations of $\Delta$v are typically of $\sim$15\%$-$20\%.  
    
\begin{figure}[!ht]
 \begin{minipage}{\linewidth}
 \begin{center}
 \includegraphics[trim = 0mm 00mm 0mm 95mm, clip, width=1\textwidth]{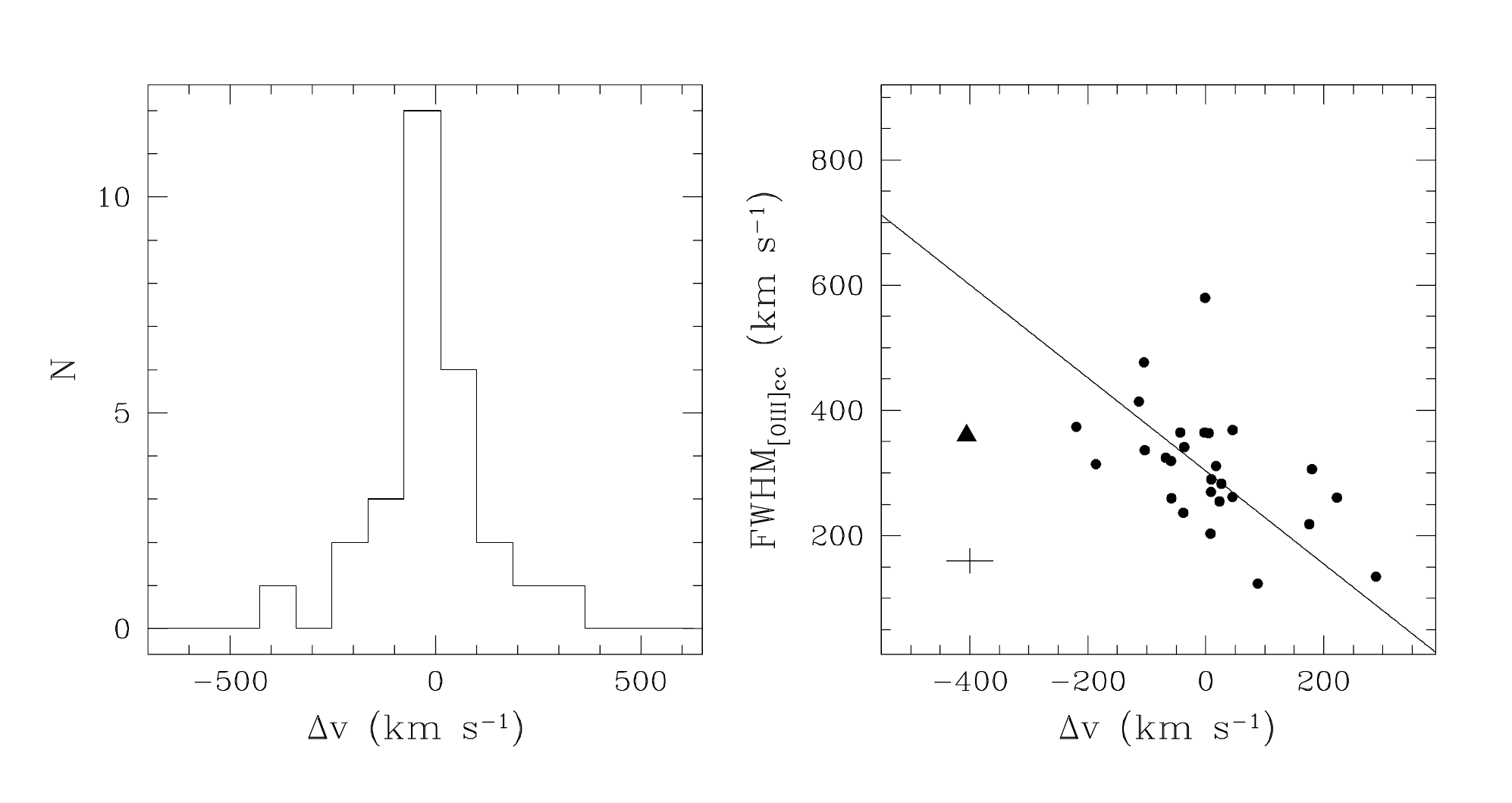}
  \end{center}
 \caption{Left: Histogram of the [OIII] blueshift relative to H$\beta$ ($\Delta$v) in units of km  s$^{-1}$. Right: The relation between the [OIII] blueshift and the FWHM of the core component of [OIII] in units of km  s$^{-1}$is shown. The blue outlier galaxy V961349$-$439 is indicated by the triangle. The typical error bar is shown. The solid line represents the best fit for our data.}
\label{fig:blueshift}
 \end{minipage}
\end{figure}

    In the left panel of Fig. \ref{fig:blueshift}, we present the distribution of the [OIII] blueshift, $\Delta$v, in units of km s$^{-1}$, which has a standard deviation of 137 km s$^{-1}$ and an interquartile range (IQR) of 104 km s$^{-1}$. According to the criteria adopted by \cite{2002ApJ...576L...9Z}, there is one blue outlier galaxy with $\Delta$v $< -$250 km s$^{-1}$. This object is V961349$-$439 and has a value of $\Delta$v of $-$405 $\pm$47 km s$^{-1}$. This object presents two blueshifted broad components of $\sim$ 3100 km s$^{-1}$ and 1300 km s$^{-1}$ and one narrow component of 360 km s$^{-1}$ in the H$\beta$ emission profile. On the other hand, the [OIII]$\lambda\lambda$4959,5007 lines of this galaxy show a core component of $\sim$ 360 km s$^{-1}$ and a blueshifted component of $\sim$ 680 km s$^{-1}$.
    
    Galaxies with larger [OIII] blueshifts tend to present broader FWHM for the core component of [OIII]. That can be seen in the right panel of Fig. \ref{fig:blueshift}, which shows the relation between $\Delta$v and the FWHM of [OIII]$_{cc}$, which have a modest Pearson coefficient correlation of r$_{p}=-$0.48 and a p-value of 9 $\times$ 10$^{-3}$. This agrees with previous results of \cite{2005MNRAS.364..187B} and \cite{2008K}. The blue outlier galaxy V961349$-$439 is indicated in the following plots as a triangle. The solid line in the plot indicates the OLS bisector fit for our data, which is 
  FWHM$_{[OIII]cc}=$(303$\pm$18) km s$^{-1}$ $-$(0.74$\pm$0.16)$\Delta$v (km s$^{-1}$).

    \subsection{Blue wings}
    \label{bluewings}
    
\begin{figure}[!ht]
 \begin{minipage}{\linewidth}
 \begin{center}
 \includegraphics[trim = 0mm 00mm 0mm 95mm, clip, width=1\textwidth]{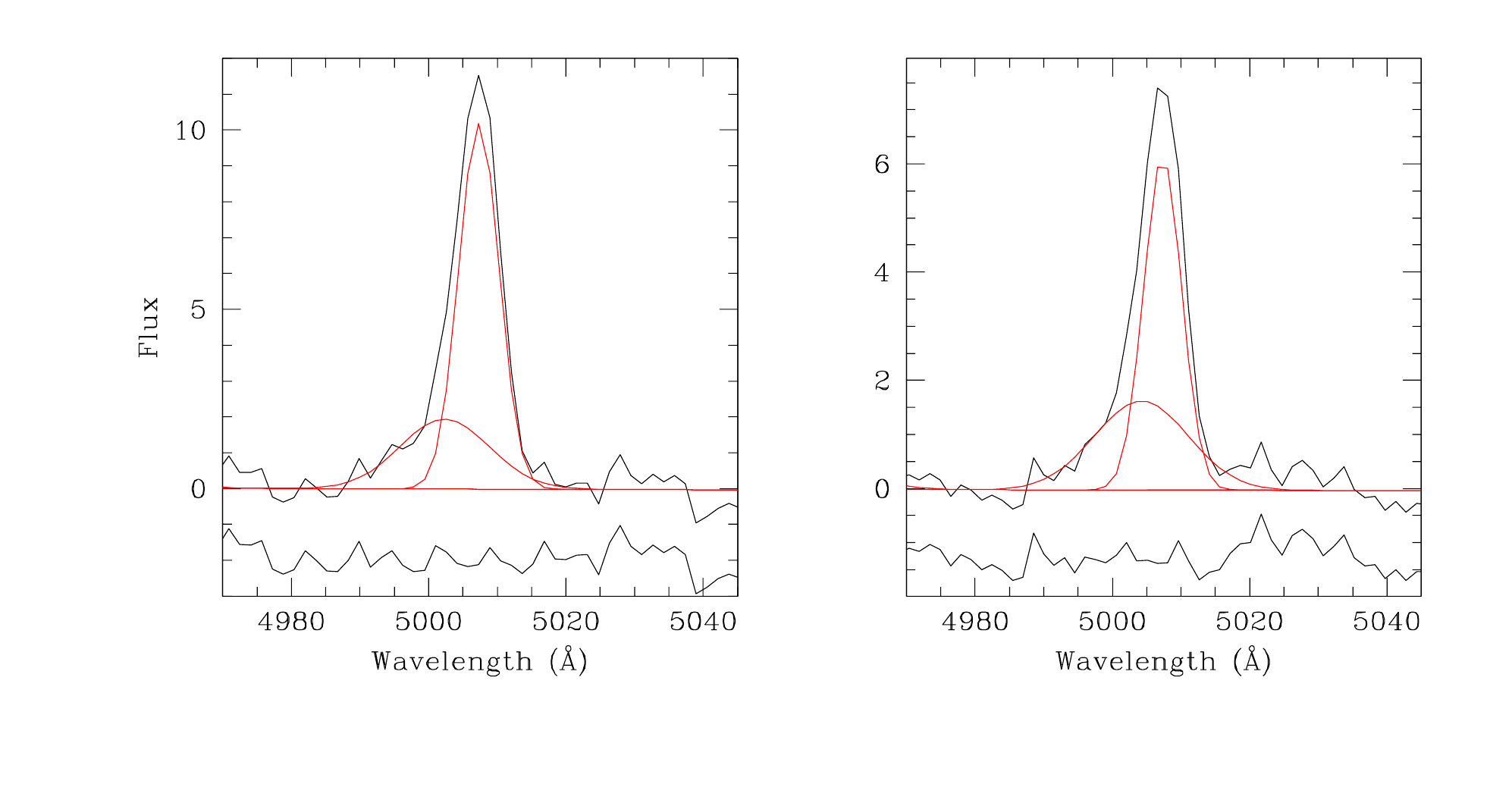}
  \end{center}
 \caption{Example of the two Gaussian fit for the [OIII]$\lambda$5007 profile of the galaxies Zw049.106 (left) and SDSSJ134524.70$-$025939.8 (right). Thick line represents the spectra; the core and asymmetric components of the [OIII] line are shown with dotted lines. The fit (sum of the two components) is superimposed on the spectra. The residuals are plotted at the bottom of each panel. The flux is in arbitrary units in both plots.}
\label{fig:oiiifit}
 \end{minipage}
\end{figure}

    As was previously mentioned, all the galaxies of the sample present asymmetries (most of which are very evident) in the [OIII]$\lambda$5007 emission profiles and were fitted with an additional Gaussian component (Sect. \ref{emissionlines}). As an example, Fig. \ref{fig:oiiifit} shows the fits of the [OIII]$\lambda$5007 line for two galaxies, where the composite profile is shown with a thick solid line, the core and asymmetric components are shown with a dotted line, and the residuals of the fit are plotted below. 
    
   We defined the parameter $\Delta$v$_{asym}$ as the difference of velocities of the centroid of [OIII]$_{ac}$ and that of [OIII]$_{cc}$. Typical uncertainties in the measurements of $\Delta$v$_{asym}$ are on the order of $\sim$15\%$-$20\%.

\begin{figure}[!ht]
 \begin{minipage}{\linewidth}
 \begin{center}
 \includegraphics[trim = 0mm 00mm 0mm 95mm, clip, width=1\textwidth]{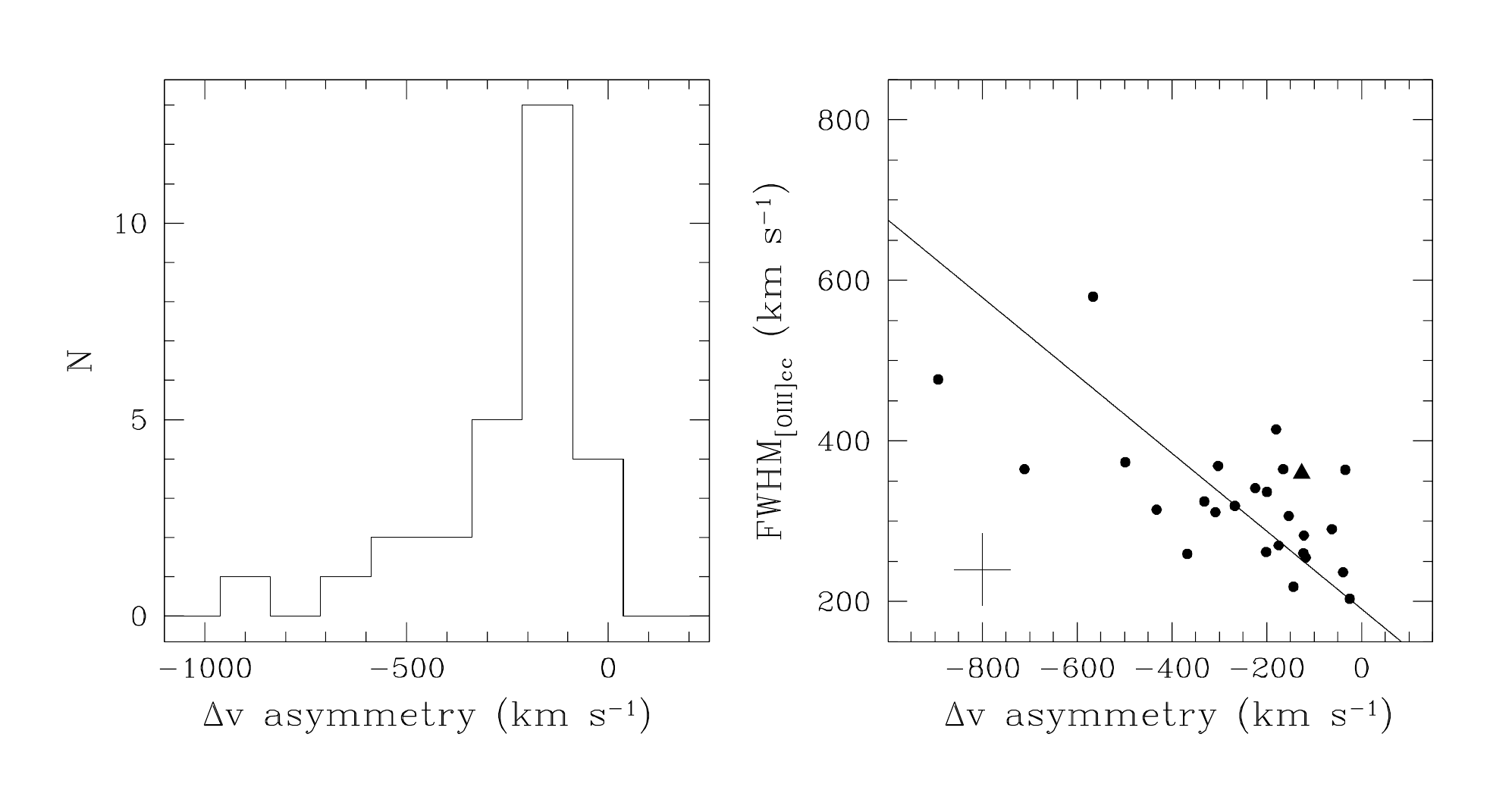}
  \end{center}
 \caption{Histogram of $\Delta$v$_{asym}$ (left). Relation between $\Delta$v$_{asym}$ and FWHM$_{[OIII]cc}$ (right) is shown. Both parameters are in units of km s$^{-1}$. The triangle indicates the blue outlier galaxy, V961349$-$439. The solid line represents the best fits for our data. A typical error bar is shown in the plot.}
\label{fig:deltaasy}
 \end{minipage}
\end{figure}

   The left panel of Fig. \ref{fig:deltaasy} shows the distribution of $\Delta$v$_{asym}$, which has a standard deviation of 207 km s$^{-1}$ and IQR of 187 km s$^{-1}$. The range of the asymmetric emission of [OIII]$\lambda$5007 is in agreement with previous known results \citep[e.g.,][]{2001A&A...372..730V}. There are some galaxies with  very marked asymmetries ($\Delta$v$_{asym}<-$250km s$^{-1}$): IRAS20520$-$2329 ($\Delta$v$_{asym}=-$267 km s$^{-1}$), Zw049.106 ($\Delta$v$_{asym}=-$302 km s$^{-1}$), CTSM02.47 ($\Delta$v$_{asym}=-$308 km s$^{-1}$), RXJ0902.5$-$0700 ($\Delta$v$_{asym}=-$ 331 km s$^{-1}$), 2MASXJ01413249$-$1528016 ($\Delta$v$_{asym}=-$367 km s$^{-1}$), WPV85007 ($\Delta$v$_{asym}=-$432 km s$^{-1}$), IRAS04576$+$0912 ($\Delta$v$_{asym}=-$498 km s$^{-1}$), RBS0219 ($\Delta$v$_{asym}=-$566 km s$^{-1}$), RXJ0024.7$+$0820 ($\Delta$v$_{asym}=-$711 km s$^{-1}$), and RXJ2301.8$-$5508 with $\Delta$v$_{asym}=-$893 km s$^{-1}$ (see Table \ref{tab:oiii}).

    The galaxies with lower values of $\Delta$v$_{asym}$, also tend to present higher FWHM of [OIII]$_{cc}$, in agreement with recent  results \citep{2016MNRAS.462.1256C}. This is shown in the right panel of Fig. \ref{fig:deltaasy}. This relation also has a modest Pearson correlation coefficient of r$_{p}=-$0.59 with a p-value $<$ 10$^{-3}$. The solid line in the plot represents the OLS bisector fit for our sample, which is FWHM$_{[OIII]cc}=$(169$\pm$29) km s$^{-1}$ $-$(0.63$\pm$0.14)$ \times \Delta$v$_{asym}$(km s$^{-1}$).

    No correlation is seen considering $\Delta$v$_{asym}$ and the FWHM of the remaining emission lines. On the other hand,  we can not say that there is a correlation between $\Delta$v$_{asym}$ and the [OIII] blueshift $\Delta$v either. The two parameters seem to be weakly correlated, given we find a Pearson correlation coefficient of r$_{p}=$0.33 and a p-value of 0.08, possibly because of the small number of objects.

    \subsection{Total asymmetry}
    \label{totalasy}
    To study the total asymmetric emission, that is, the total extension of the blue wing, in addition to $\Delta$v$_{asym}$ we have to consider half of the width of the asymmetric component in the base, i.e.,
    \begin{equation}
    total \ asymmetry = \Delta v_{asym} - \frac{1}{2} Width_{base}[OIII]_{ac}
    .\end{equation}
    
  Although it is difficult to estimate the width of the base, we can assume that Width$_{base}$[OIII]$_{ac}$ is $\sim$ 2 FWHM[OIII]$_{ac}$, then we have
    
    \begin{equation}
\label{eq:totalasym}
total \ asymmetry =  \Delta v_{asym} - FWHM[OIII]_{ac}
.\end{equation}
    
    In this way, the thus defined total asymmetry parameter is a good estimator of the radial velocity of the extreme of the blue wing relative to the centroid of the core component of [OIII]$\lambda$5007. 
    As mentioned in Sect \ref{emissionlines}, to estimate the uncertainties of the measurements, we repeated the measurement procedure several times in galaxies with different S/N. In this manner, the relative error of the total asymmetry is typically 4\%$-$15\%.
    
       \begin{figure}[!ht]
 \begin{minipage}{\linewidth}
 \begin{center}
 \includegraphics[trim = 0mm 00mm 0mm 95mm, clip, width=1\textwidth]{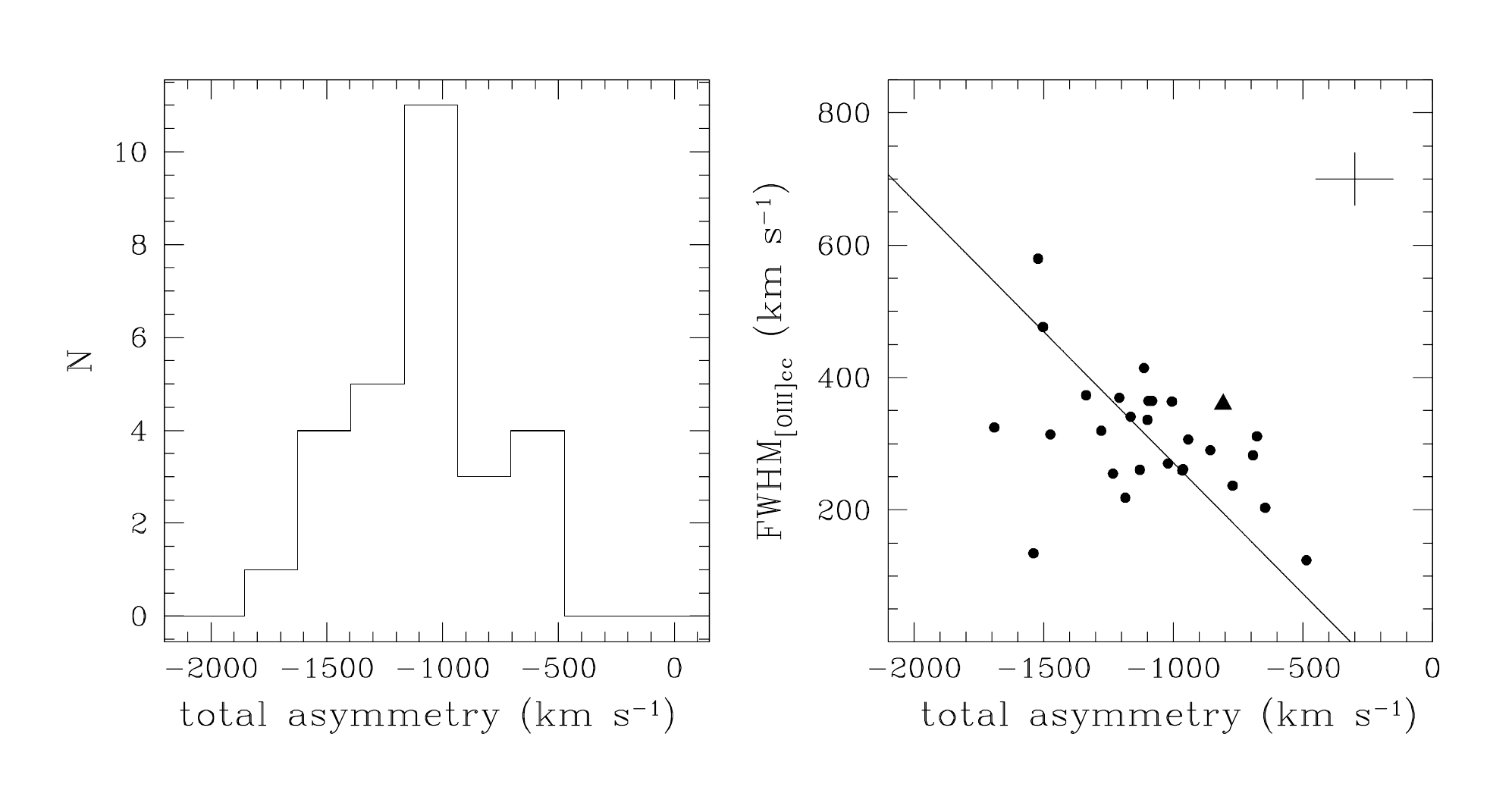}
  \end{center}
 \caption{Histogram of the total asymmetry, given by equation \ref{eq:totalasym} (left). The relation between the total asymmetry and FWHM$_{[OIII]cc}$ (right) is shown. Both parameters are in units of km s$^{-1}$. The triangle indicates the blue outlier galaxy, V961349$-$439. The solid line represents the best fit for our data. A typical error bar is shown.}
\label{fig:asyt}
 \end{minipage}
\end{figure}

   \begin{figure*}[!ht]
 \begin{minipage}{\linewidth}
 \begin{center}
 \includegraphics[trim = 0mm 00mm 0mm 132mm, clip, width=1\textwidth]{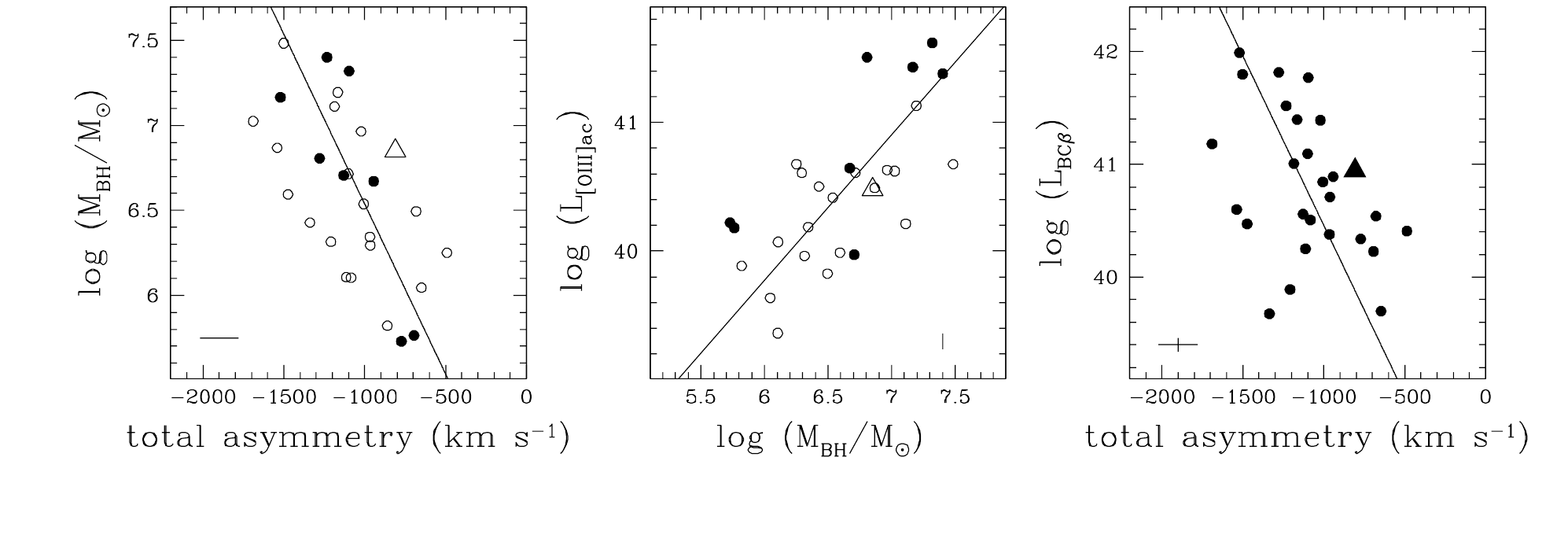}
  \end{center}
 \caption{Left: Relation between the total asymmetry (given by equation \ref{eq:totalasym}), in units of km s$^{-1}$,  and the black hole mass (in units of M$_{\odot}$). Middle: The correlation among the black hole mass and the luminosity of the blue wing, [OIII]$_{ac}$ (in units of erg s$^{-1}$) is shown. Right: The tendency between the total asymmetry (in km s$^{-1}$) and the luminosity of the broad component of H$\beta$ (in units of erg s$^{-1}$) is shown. Open symbols represent galaxies whose black hole mass estimations were extracted from \cite{Schmidt}. The triangle indicates the blue outlier galaxy V961349$-$439. The solid lines represent the best fit for our data. Typical error bars are shown.}
\label{fig:oasy}
 \end{minipage}
\end{figure*}
    
    The left panel of Fig. \ref{fig:asyt} shows the distribution of the total asymmetry, in units of km s$^{-1}$. The galaxies of our sample have [OIII] blue wings with an extension (relative to the centroid of [OIII]$_{cc}$) that ranges from $\sim -$1700 km s$^{-1}$ to $\sim-$500 km s$^{-1}$;  most galaxies show blue wings of $\sim -$1100 km s$^{-1}$. The distribution has a standard deviation of 298 km s$^{-1}$ and IQR of 375 km s$^{-1}$. Just like $\Delta$v and $\Delta$ v$_{asym}$, the total blue wing also correlates with the FWHM of [OIII]$_{cc}$ (right panel). This tendency has a modest Pearson correlation coefficient of r$_{p}=-$0.46 and a p-value of 0.01. The solid line represents the OLS bisector fit for our data, which is FWHM$_{[OIII]cc}=$(-125$\pm$113)km s$^{-1}$ $-$(0.40$\pm$0.09) $\times$  total asymmetry (km s$^{-1}$).

     On the other hand, to study the relation between the blue wing and central engine, it is necessary to estimate the black hole mass. For the 28 galaxies of our sample, 20 of these sources already have a black hole mass estimation in \cite{Schmidt}. We calculated the black hole mass of the 8 remaining galaxies, using equation 6 of \cite{greene}. 
     For 6 of the galaxies, we used the FWHM and luminosity of the broad component of H$\alpha$ because the black hole mass estimation through this line is more reliable than the estimation through H$\beta$ because of its higher S/N ratio \citep{greene}. For the 2 remaining galaxies, the parameters of the broad component of H$\beta$ were used through equation 7 of the mentioned work.\\ 
     The used formalism in the black hole estimation assumes the correlation between the size of the BLR and the luminosity at 5100 \AA \ \citep[see][]{greene}. According to some authors, for highly accreting sources, this scaling relation seems to deviate from those of lower accretion rate \citep[e.g.,][]{2014ApJ...793..108W,2014ApJ...792L..13W}. Related to this, for a given L$_{5100}$, a high accretion source has shorter H$\beta$ lags by a factor of up to $\sim$ 3 $-$ 4 compared with lower accretion rate objects \citep{2015ApJ...806...22D}, indicating a smaller size of the BLR. This difference in the estimation of the BLR size for high accreting galaxies produces a black hole mass uncertainty of 0.5 dex.\\     
     Table \ref{tab:mass1} presents the FWHM and luminosity of BC$\alpha$ and the black hole masses for 8 galaxies. They are in the same range of black hole mass for NLS1 galaxies previously studied \citep[e.g.,][]{2004AJ....127.1799G,Schmidt,2017A&A...606A...9J}.

    Analyzing the total asymmetry, we found a very interesting correlation between this asymmetry and the black hole mass; we find  a Pearson coefficient of r$_{p}=-$0.63 and a p-value of 3 $\times$10$^{-4}$. This tendency is stronger than the correlation between M$_{BH}$ and FWHM$_{[OIII]ac}$ (r$_{p}=$0.53) and that between M$_{BH}$ and $\Delta$v$_{asym}$ (r$_{p}=-$0.20). Considering galaxies with blue asymmetries in the [OIII]$\lambda$5007 profile, objects with higher black hole masses tend to show more extended blue wings (left panel of Fig. \ref{fig:oasy}). 
    
   On the other hand, the black hole mass also correlates with the luminosity of [OIII]$_{ac}$ (middle panel). This tendency has a moderate Pearson correlation coefficient of  r$_{p}=-$0.68 and a p-value of 7 $\times$ 10$^{-5}$. Galaxies with larger black hole masses seem to present higher luminosities of the blue wing. The solid line in the plot represents the OLS bisector fit for our data, which is log($L_{[OIII]ac}$)$=$(33$\pm$1)$+$(1.13$\pm$0.15) $\times$ log(M$_{BH}$/M$_{\odot}$).

   If the total asymmetry correlates with the central black hole mass, we can also expect a correlation between the former and the luminosity of the broad component of H$\beta$. This tendency can be seen in the right panel of Fig. \ref{fig:oasy}, where the solid line represents the best OLS bisector fit for the data. This relation has a Pearson coeficient of r$_{p}=-$0.43 and a p-value < 0.03.

    The Eddington luminosity, given by $L_{Edd}=1.26 \times 10^{38} M_{BH}/M_{\odot} \ ergs^{-1}$, is closely linked to black hole mass.  We calculated the Eddington ratio between the bolometric luminosity (L$_{bol}$) and the Eddington luminosity, considering $L_{bol} \sim 3500 \times L_{[OIII]}$ \citep{2004ApJ...613..109H}.
    Assuming that the uncertainty of the Eddington ratio is given by the error of the luminosity of [OIII] and the uncertainty of the black hole mass estimation, the resulting lower limit for the error of the Eddington ratio is $\sim$0.5 dex.\\  
As expected, the luminosities of some objects in our sample are beyond the Eddington limit, which is in agreement with previous results \citep[e.g.,][]{2016MNRAS.462.1256C,Xu2007,Bian2004}. 
     Considering the Eddington ratio and $\Delta$v, no significant correlation was found between both parameters, with a Pearson correlation coefficient of r$_{p}=-$0.13 and a p-value of 0.5; this agrees with results of \cite{2016MNRAS.462.1256C} and \cite{2005MNRAS.364..187B}. In addition, given a Pearson coefficient of
r$_{p}=-$0.09, no correlation was found between the Eddington ratio and $\Delta$v$_{asym}$.
     On the other hand, a weak correlation was found between the Eddington ratio and the total asymmetry, with the  Pearson coefficient of r$_{p}=$0.33 and a p-value of 0.08. Also, an anti-correlation between the Eddington ratio and the black hole mass is observed and the correlation coefficient of r$_{p}=-$0.42 and p-value$<$0.03.
   
   Table \ref{tab:oiii} presents the FWHM and the flux of the core and the asymmetric emission of [OIII]$\lambda$5007, the values of the [OIII] blueshift ($\Delta$v), the amount of asymmetry ($\Delta$v$_{asym}$), the total asymmetry (equation \ref{eq:totalasym}) and the Eddington ratio. All presented FWHM were corrected by the instrumental broadening, as was mentioned in Sect. \ref{emissionlines}.

    As we said in Sect. \ref{observation}, 20 of the 28 studied objects in this work were taken from the sample of 36 galaxies of \cite{Schmidt} because they present asymmetric [OIII] profiles. The black hole mass range of galaxies that show asymmetric profiles is almost the same as that of galaxies that show symmetric lines. This means that blue wings can be present in different galaxies, independently of the black hole mass. In this scenario, galaxies with a large black hole mass may show non-asymmetric [OIII] emission profiles and galaxies with lower black hole masses may exhibit asymmetric lines. Related to this, 94\% of the sample of NLS1 of \cite{2016MNRAS.462.1256C} show asymmetric [OIII] emission profiles, and these authors have claimed that no difference appears to exist between NLS1 and BLS1, despite the difference in the black hole masses. In order to try to detect any possible difference between galaxies that show asymmetric profiles and objects that do not, we checked the Eddington ratio. Both groups of galaxies present almost the same range of Eddington ratio.
    
    Although we do not know which mechanisms trigger the outflows, when we consider only galaxies with asymmetric [OIII] emission profiles, we found that there is a correlation between the total asymmetry and the black hole mass. Nevertheless, there is no correlation between these parameters when considering a sample that includes symmetric emitting galaxies.

   \section{Discussion}
   \label{sec:discussion}

   In a study of the presence of winds in a large spectroscopic sample of around 600000 local galaxies, \cite{2017A&A...606A..36C}  found that in star forming galaxies, the ionized interstellar gas traced by the [OIII]$\lambda$5007 line never appears to be outflowing. In these galaxies, the [OIII] line profile is perfectly fitted by a single Gaussian without need of a second component. These authors claimed that the need for a second broader Gaussian component increases with a clear trend with the increase of the AGN contribution to the galaxy spectrum. In this scenario, blue wings were found in 73\% of a sample of intermediate-type Seyfert and in 68\% of a sample of Seyfert 2 galaxies \citep{2012MNRAS.427.1266V}. \cite{2016ApJ...817..108W} studied the outflows in a sample of type 2 AGNs and they found that 43\% of the galaxies present winds. These decreasing percentages,
going from type 1 to type 2 AGNs, are in agreement with the results of \cite{2017A&A...606A..36C}. This is in compliance with the fact that asymmetric profiles are usually detected in quasars, Seyfert galaxies, and in particular are very common in NLS1 \citep[e.g.,][]{2001A&A...372..730V,2011ApJ...739...28X}. In a sample of 296 NLS1, \cite{2016MNRAS.462.1256C} found that 94\% of the galaxies exhibit [OIII] emission profiles with asymmetries.
   
   In our sample of 28 asymmetric-line emitting NLS1 galaxies, the two components of the [OIII]$\lambda$5007 profile show different line widths and have velocities that range between 200 and 800 km s$^{-1}$ for the core component and between 300 and 1400 km s$^{-1}$ for the asymmetric component. This can be interpreted as two kinematically distinct regions, which in agreement with some authors \cite[e.g.,][]{2003MNRAS.342..227H,2005MNRAS.364..187B}. The range of velocities of the asymmetric component of [OIII] has values between the one of the core component and the range of the BLR (Sect. \ref{emissionlines}). If we assume that the square of the FWHM of the gas decreases with the distance to the center \citep[e.g.,][]{Schmidt}, one could expect that the asymmetric component is originated at the inner parts of the NLR between the emitting region of the core component of [OIII]$\lambda$5007 line and the BLR (by definition, the emitting region of the broad component of H$\beta$). This indicates that the kinematic of the gas in the inner parts of the NLR is more turbulent compared to the external emitting regions, in agreement with several studies \citep[e.g.,][]{2001A&A...372..730V,2005MNRAS.364..187B}. Related to this, \cite{2016MNRAS.462.1256C} suggested that the asymmetry is probably caused
by the presence of outflowing gas from the inner regions of the active nucleus, which interacts with the surrounding medium by transferring kinetic energy and by reducing the equivalent width of the [OIII] line as its velocity increases. This is in agreement with the results of \cite{2012ApJ...756...51L}, who found that the shift of the blue wing correlates with the equivalent width of the [OIII] emission.

        The blue outlier galaxy V961349$-$439 does not deviate in any of the studied relations. These tendencies mainly involve parameters of the blue wing such as the total asymmetry, the luminosity of the asymmetric component, and $\Delta$v$_{asym}$. There is no significant correlation between the mentioned parameters and the blueshift $\Delta$v as the Pearson coefficient values are between $\sim$0.1 and $\sim$0.3. This is in agreement with the idea that the [OIII] blueshift is not connected to the blue wing \citep{2005AJ....130..381B, 2016MNRAS.462.1256C}. This could suggest that the mechanism that produces the blueshift is not the same as that yielding the asymmetry of the profile. 
    
    Considering the case of V961349$-$439, this object presents a higher [OIII] blueshift relative to H$\beta$ ($\Delta$v$=-$405 km s$^{-1}$), but shows a blue wing with an extension of $\sim -$810 km s$^{-1}$, which is one of the lower values of the sample; this is in concordance with the analysis above.

  However, it has been proposed that the blueshift of [OIII] is the result of an outflowing gas from the central regions \citep{2002ApJ...576L...9Z}. The outflowing gas could be originated at the inner NLR and is possibly related to the wind of the nucleus \citep{2000ApJ...545...63E}. In this scenario, both the blueshift and the blue wings would be originated at the same region. 
  
  Aside from the [OIII] emission, there are high ionization emission lines emitted at the NLR, which also show shifts, as is the case of [Fe VII] \citep{RA2006,2016MNRAS.462.1256C}, [Ne III] and [Ne V] \citep{2009ApJ...702L..42S}, and [Fe VII], [Fe X] and [Fe XI] \citep{RA2006,2008K}. Moreover, the high ionization C IV$\lambda$1549 line, emitted from the BLR, was also found to be blueshifted \citep[e.g.,][]{2007ApJ...666..757S,2007AIPC..938..104K}. In this scenario, and considering that the higher the ionization potential, the line originates in more internal regions, the outflow would be moving through a stratified medium ionized by the central engine \citep[e.g.,][]{2016MNRAS.462.1256C}.

    On the other hand, the found correlation between the black hole mass and the total asymmetry, and the relation between the former and the luminosity of the asymmetric component of [OIII] line, suggest that the winds originated at the NLR and manifested through the [OIII]$_{ac}$ emission are affected by the central engine. This would be in agreement with the fact that the black hole mass correlates with the radio emission, suggesting that NLS1 with larger black hole masses are more likely to show relativistic jets \citep{2015A&A...573A..76J} and therefore, strong winds. In addition, the correlation between the total asymmetry and the black hole mass is in agreement with the idea that wider extended components could be explained by virial effects of the central black holes, as suggested by \cite{2017MNRAS.468..620Z}.
    
    We found that the Eddington ratios of the galaxies of the sample are in the same range as the obtained by other authors \citep{Xu2007,Bian2004}; some objects show luminosities close to the Eddington luminosity, as it is expected in NLS1 galaxies \citep[e.g.,][]{2016MNRAS.462.1256C,Xu2007}.
    
    There is a mild correlation between the total asymmetry and the Eddington ratio, for which we find a Pearson correlation coefficient of r$_{p}=-$0.33. We also found trends between the total asymmetry and the luminosities of the broad component of H$\beta$ and [OIII]$\lambda$5007 with r$_{p}=-$0.43 and r$_{p}=-$0.40, respectively. The fact that the total asymmetry correlates with the Eddington ratio, the emission of the BLR, and the emission of [OIII] could indicate that the extension of the blue wing also depends on the intensity of the AGN; this is in agreement with previous results of \cite{2013MNRAS.433..622M} and \cite{2016ApJ...817..108W} in type 2 AGNs. Nevertheless, these correlations are not so marked as the relation between the total asymmetry and the black hole mass, which has a Pearson correlation coefficient of r$_{p}=-$0.63. This implies, as is expected, that asymmetric profiles are more related to kinematical processes than to photoionization mechanisms.
   
   Related to the [OIII] profile, some studies found that the bulge of the host galaxy has influence in the width of the emission lines emitted at the NLR, i.e., the [OIII]$\lambda$5007 line \citep[e.g.,][]{1996ApJ...465...96N}. As we showed, the asymmetric emission of [OIII] has a marked relation with the central engine, not only in the shape, but also in the luminosity. Thus, it could be said that the [OIII]$\lambda$5007 emission profile is doubly affected, by the bulge of the host galaxy and by the central engine of the AGN.

    \section{Final remarks}
    \label{final}
    
    We analyzed a sample of 28 poorly studied NLS1 galaxies, most of which are from the southern hemisphere. All of these galaxies present a blue asymmetric component in the [OIII]$\lambda$5007 emission line. 
    After a careful Gaussian decomposition, we measured the radial velocity difference between the core component of [OIII]$\lambda$5007 and the narrow component of H$\beta$. We found that many of the galaxies present a blueshift of [OIII] relative to H$\beta$. The distribution of $\Delta$v in our sample has a standard deviation of 137 km s$^{-1}$ and an IQR of 104 km s$^{-1}$. Considering the criterion adopted by \cite{2002ApJ...576L...9Z}, we found one blue outlier galaxy with a blueshift $< -$250 km s$^{-1}$. This object is V961349$-$439 and has  a blueshift of $-$405 $\pm$47 km s$^{-1}$.
    We confirmed a previously known correlation between the [OIII] blueshift relative to H$\beta$ and the FWHM of the core component of [OIII]$\lambda$5007 \citep{2005MNRAS.364..187B,2008K}, which for our sample has a Pearson correlation coefficient of r$_{p}=-$0.48 and a p-value of 9 $\times$ 10$^{-3}$. Galaxies with larger blueshift tend to present broader FWHM in the core component of [OIII].
    
    Aside from the [OIII] blueshift relative to H$\beta$, we also analyzed the blue wings of the asymmetric emission of the [OIII]$\lambda$5007 emission profile. We studied the radial velocity difference between the centroid of the asymmetric component relative to the core component, that is $\Delta$v$_{asym}$. We found that the distribution of $\Delta$v$_{asym}$ has a standard deviation of 207 km s$^{-1}$, an IQR of 187 km s$^{-1}$, and a mean value of $-$248 km s$^{-1}$. The range of $\Delta$v$_{asym}$ found in this work is in agreement with previous results \citep[e.g.,][]{2001A&A...372..730V,2016MNRAS.462.1256C}. The FWHM of the core component of [OIII]$\lambda$5007 also correlates with $\Delta$v$_{asym}$, given a Pearson correlation coefficient of r$_{p}=-$0.59 and a p-value $<$ 10$^{-3}$. This is in agreement with recent results of \cite{2016MNRAS.462.1256C}, who found that these parameters are related with a Spearman coefficient of r$_{s}=-$0.46 and a p-value of 1.4 $\times$ 10$^{-18}$.
     Considering $\Delta$v$_{asym}$ and the [OIII] blueshift $\Delta$v, a very weak tendency seems to relate both parameters; we find a Pearson correlation coefficient of r$_{p}=$0.33 and a p-value of 0.08, which is probably due to the small number of galaxies.
    
    Finally. we studied the radial velocity difference between the blue end of the blue wing and the centroid of the core component of [OIII]$\lambda$5007. In our sample, this total asymmetry (defined by equation \ref{eq:totalasym}) ranges from $\sim -$1700 km s$^{-1}$ to $\sim-$500 km s$^{-1}$, with most galaxies showing blue wings which extend up to $\sim -$1100 km s$^{-1}$. This distribution has a standard deviation of 298 km s$^{-1}$ and an IQR of 375 km s$^{-1}$.
  The total blue wing also correlates with the FWHM of [OIII]$_{cc}$, just like $\Delta$v and $\Delta$ v$_{asym}$. This tendency has a Pearson correlation coefficient of r$_{p}=-$0.46 and a p-value of 0.01. 
  Given a Pearson correlation coefficient of r$_{p}=-$0.63 and a p-value of 3 $\times$10$^{-4}$, we found that the radial velocity of the blue end of the asymmetry relative
to the core component of [OIII] correlates with the black hole mass. Galaxies with higher black hole masses tend to show a more marked asymmetry in the [OIII]$\lambda$5007 emission profile, which agrees with the idea of that NLS1 galaxies with larger black hole masses have higher probabilities of showing relativistic jets, and therefore, strong winds \citep{2015A&A...573A..76J}.\\

  We calculated the Eddington ratio for the whole sample. As expected, the luminosities of some galaxies are beyond the Eddington limit, in agreement with previous results \citep[e.g.,][]{2016MNRAS.462.1256C,Xu2007,Bian2004}. We see a weak correlation between the Eddington ratio and the total asymmetry, with a Pearson correlation coefficient of r$_{p}=$0.33 and a p-value of 0.08.\\   
  
  We found a correlation between the luminosity of the asymmetric component of [OIII] and the black hole mass, which has a moderate Pearson correlation coefficient of r$_{p}=-$0.68 and a p-value of 7 $\times$ 10$^{-5}$. Related to this, galaxies with larger black hole masses seem to show higher luminosities of the blue wing.
  Finally, we found a correlation between the luminosity of the broad component of H$\beta$ and the total asymmetry, with a Pearson coefficient of r$_{p}=-$0.43 and a p-value < 3 $\times$ 10$^{-2}$. This suggests that the blue wing is influenced by the intensity of the AGN, which is in agreement with previous results \citep{2013MNRAS.433..622M,2016ApJ...817..108W}.

\begin{acknowledgements}
      E.O.S and G.A.O want to thank the considerable help of Damian Mast and Jose Ovejero. We also appreciate the helpful comments and suggestions made by the anonymous referee, which improved this article. This work was partially supported by Consejo de Investigaciones Cient\'ificas y T\'ecnicas (CONICET) and Secretar\'ia de Ciencia y T\'ecnica de la Universidad Nacional de C\'ordoba (SecyT). This research has made use of the NASA/IPAC Extragalactic Database (NED) which is operated by the Jet Propulsion Laboratory, California Institute of Technology, under contract with the National Aeronautics and Space Administration. 
\end{acknowledgements}

\bibliographystyle{aa}
\bibliography{Bibliography}

\appendix
\section{Additional data}
\begin{figure*}[!ht]
\includegraphics[trim = 65mm 80mm 0mm 10mm, clip, width=\linewidth]{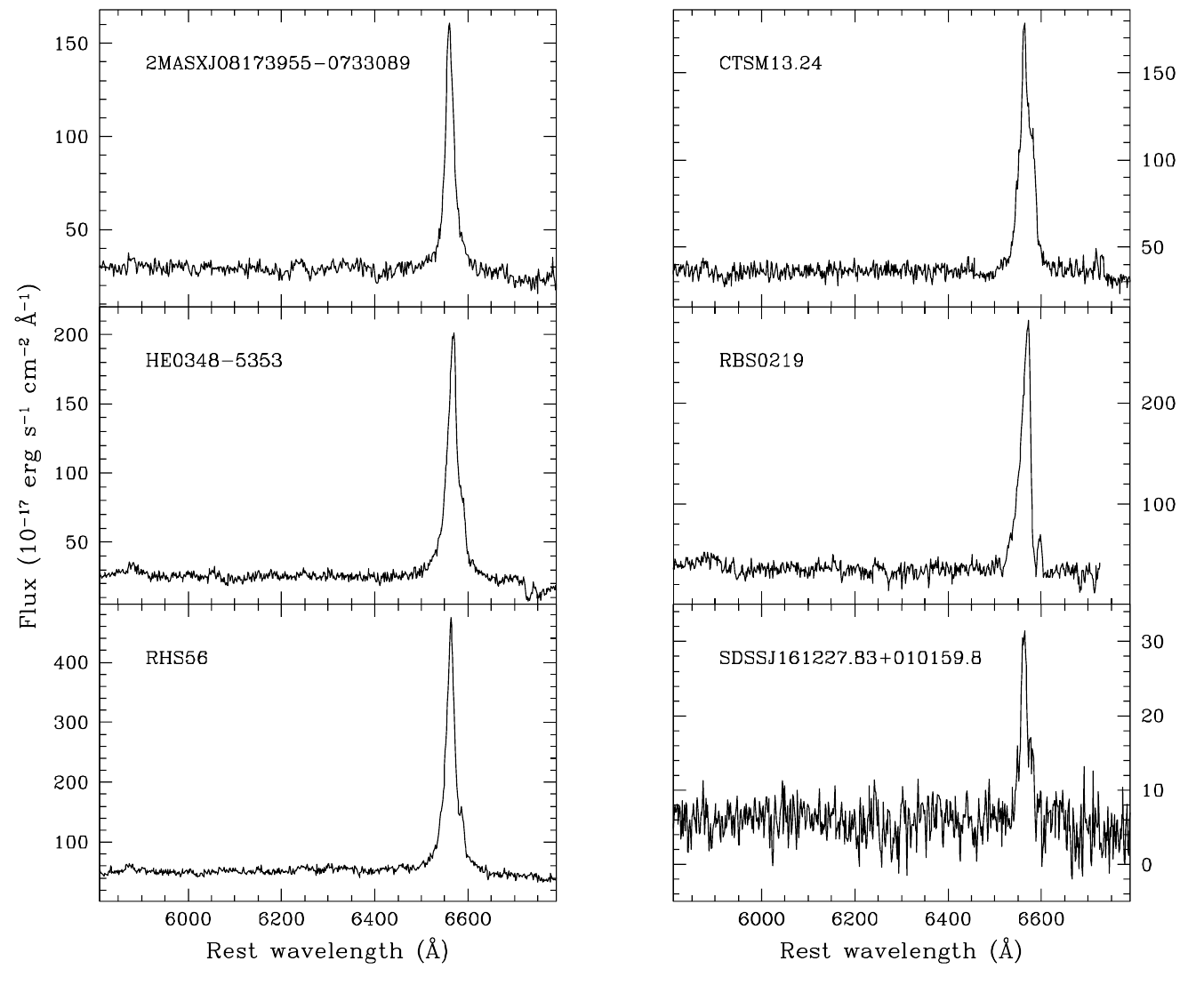}
\caption{Observed spectra of the NLS1 in the range 5800\AA{} - 6800\AA{}. Fluxes are in units of 10 $^{-17}$ erg cm $^{-2}$ s$^{-1}$ \AA{}$^{-1}$.}
\label{fig:sper}
\end{figure*}

In this Appendix, we present the principal characteristic of the galaxies of the sample (Table \ref{tab:sample}), the observational data (Table \ref{tab:obs}), black hole masses for eight galaxies (Table \ref{tab:mass1}), and the parameters of [OIII]$\lambda$5007 (Table \ref{tab:oiii}). We also show the blue spectra of the galaxies in Fig. \ref{fig:spec1} and the red spectra of six galaxies in Fig. \ref{fig:sper}.

\begin{table*}[!ht]
\caption{Sample. Column 1: Galaxy name. Columns 2 to 8: Right ascension (J2000), declination (J2000), radial velocity, redshift, apparent magnitude and filter, major diameter, minor diameter, as taken from NED. Galaxies labeled with "*" were taken from the sample of \cite{Schmidt}.}             
\label{tab:sample}      
\centering          
\begin{tabular}{l c c c c c c c }     
\hline\hline       
                    
                               &      RA      &     Dec      &  Radial vel.  &            &               & Major Diam &  Minor Diam \\
  Galaxy name                  &   J(2000.0)  &   J(2000.0)  &  (km s$^{-1}$)      &  Redshift  &  Mag \& Filter& (arc min)   &  (arc min)   \\
\hline                    
 1RXSJ040443.5$-$295316*           &  04 04 43.28  &  $-$29 53 23.1  &  17995  &   0.060025 &   17.3V &       0.20 &       0.19 \\ 
2MASXJ01413249$-$1528016*         &  01 41 32.50  &  $-$15 28 01.5  &  24304  &   0.081069 &  16.9V &       0.34 &       0.24 \\
2MASXJ05014863$-$2253232*        &  05 01 48.63  &  $-$22 53 23.2  &  12232  &   0.040800 &   14.0V &       0.31 &       0.29 \\
2MASXJ08173955$-$0733089         &  08 17 39.58  &  $-$07 33 08.8  &  21765  &   0.072600 &  16.9R &       ---  &       --- \\    
2MASXJ21531910$-$1514111*         &  21 53 19.13  &  $-$15 14 12.1  &  23324  &   0.077800 &   14.7V &       ---  &       ---   \\      
2MASXJ21565663$-$1139314*         &  21 56 56.61  &  $-$11 39 30.6  &   8420  &   0.028086 &  15.4V &       ---  &       ---  \\      
CTSH34.06                        &  06 09 17.48  &  $-$56 06 58.4  &   9535  &   0.031805 &   16.2V &       0.45 &       ---   \\   
CTSM02.47*                        &  10 46 23.55  &  $-$30 04 21.0  &  17106  &   0.057060 &     17.0V &       0.18 &        0.2 \\
CTSM13.24                        &  13 12 28.60  &  $-$25 17 00.7  &  19025  &   0.063460 &   17.5V &       0.18 &        --- \\  
FAIRALL0107*                      &  21 35 29.50  &  $-$62 30 07.2  &  18275  &   0.060959 &  16.7b &       0.44 &       0.34 \\
HE0348$-$5353                    &  03 49 28.00  &  $-$53 44 47.0  & $>$30000&   0.130000 &   15.7V &       ---  &        --- \\  
HE1107$+$0129*                    &  11 10 12.07  &  $+$01 13 27.8  &  28630  &   0.095500 &  16.5g &       0.31 &       0.30 \\
IRAS04576$+$0912*                 &  05 00 20.77  &  $+$09 16 55.6  &  10822  &   0.036098 &   16.6V &       ---  &        --- \\  
IRAS16355$-$2049*                 &  16 38 30.92  &  $-$20 55 24.6  &   7906  &   0.026372 &   14.5V &       0.30 &       ---  \\  
IRAS20520$-$2329                 &  20 54 57.30  &  $-$23 18 24.0  & $>$30000&   0.206300 &  16.3R &       ---  &       ---  \\  
MCG$-$04.24.017*                  &  10 05 55.37  &  $-$23 03 25.0  &   3842  &   0.012816 &  14.5V &        1.0 &        0.5 \\
RBS0219                          &  01 35 27.00  &  $-$04 26 35.0  & $>$30000&   0.154700 &   16.5V &       0.23 &       0.20  \\
RHS56                            &  20 39 27.19  &  $-$30 18 52.2  &  23709  &   0.079085 &   16.0V &       ---  &       ---  \\  
RXJ0024.7$+$0820*                 &  00 24 45.70  &  $+$08 20 56.9  &  20086  &   0.067000 &   18.2V &       ---  &       ---  \\  
RXJ0323.2$-$4931*                 &  03 23 15.35  &  $-$49 31 06.7  &  21285  &   0.071000 &  12.4R &       0.48 &       0.30 \\ 
RXJ0902.5$-$0700*                 &  09 02 33.57  &  $-$07 00 04.3  &  26715  &   0.089112 &  17.7b &       ---  &       ---  \\  
RXJ2301.8$-$5508*                 &  23 01 52.01  &  $-$55 08 31.1  & $>$30000&   0.141000 &  14.7R &       ---  &       ---  \\  
SDSSJ134524.70$-$025939.8*        &  13 45 24.70  &  $-$02 59 39.8  &  25566  &   0.085279 &  16.8g &       0.27 &       0.17 \\
SDSSJ161227.83$+$010159.8        &  16 12 27.84  &  $+$01 01 59.9  &  29109  &   0.097096 &  18.6g &       ---  &       ---  \\  
SDSSJ225452.22$+$004631.4*        &  22 54 52.22  &  $+$00 46 31.4  &  27202  &   0.090735 &  17.3g &       ---  &       ---  \\  
V961349$-$439*                    &  13 52 59.63  &  $-$44 13 25.4  &  15589  &   0.052000 &   15.4V &       ---  &       ---  \\  
WPV85007*                         &  00 39 15.85  &  $-$51 17 01.5  &   8577  &   0.028610 &  15.8V &       0.18 &       0.15 \\
Zw049.106*                        &  15 17 51.70  &  $+$05 06 27.7  &  11626  &   0.038780 &   15.6V &       0.70 &       0.43 \\
\hline                  
\end{tabular}
\end{table*}

\begin{table*}
\caption{Observed galaxies. Column 1: Galaxy name. Column 2: Observation date (mm/dd/yyyy) of the blue range spectra ($\sim$ 4300 \AA{} $-$ $\sim$ 5200 \AA{}). Column 3: Exposure time of the blue range. Column 4: Observation date (mm/dd/yyyy) of the red range spectra ($\sim$ 5800 \AA{} $-$ $\sim$ 6800 \AA{}). Column 5: Exposure time of the red range. Column 6: S/N in the range 4750 \AA{} $-$ 4800 \AA{}. The red spectra of galaxies labeled with "*" are available in \cite{Schmidt}. The S/N of the object labeled with "$\dagger$" was measured in the spectral range 5070 \AA{} $-$ 5120 \AA{}.}         
\label{tab:obs}      
\centering          
\begin{tabular}{l c c c c c}     
\hline\hline       

                         & Date         & Exposure time     & Date           & Exposure time   & S/N     \\
Galaxy name                  & blue range       &   (s)         & red range      & (s)  & (4750 \AA{}$-$ 4800 \AA{}) \\  \hline

1RXS J040443.5$-$295316*    & 10/26/2014         &      3$\times$2400  & ---  &  ---     & 10    \\
2MASX J01413249$-$1528016*  & 10/28/2014 &      3$\times$2400  & ---   &        ---  & 12   \\
2MASXJ05014863$-$2253232*   & 10/13/2015 &      3$\times$2400  & ---    &       --- &  21   \\
2MASXJ08173955$-$0733089   & 04/12/2013 &       3$\times$2400  &  04/10/2013  &      3$\times$1800  &  7             \\
2MASX J21531910$-$1514111*  & 10/26/2014 &      2$\times$2400  & ---  & ---  &  11          \\
2MASXJ21565663$-$1139314*   & 10/16/2015 &      3$\times$2400  & ---   &        ---       & 21   \\
CTSH34.06                 & 10/19/2015  &       4$\times$2400  &  ----        &  ----    &  22   \\
CTSM02.47*                & 04/26/2012  &       3$\times$1800  &  ---   &---    &  6 \\
CTSM13.24               & 04/29/2014    &       3$\times$2400  & 04/10/2013     &3$\times$1800  &  13     \\      
FAIRALL0107*              & 10/13/2015 &        4$\times$1800  & ---   &        ---       &       24 \\
HE0348$-$5353           & 10/28/2014 &          3$\times$2400  & 10/27/2014   &     3$\times$2400    &  11          \\
HE1107$+$0129*           & 04/08/2013 &         3$\times$1800  & ---   &        ---       &        17  \\
IRAS04576$+$0912*        & 10/18/2015 &                 3$\times$2400  & ---   & ---   & 21      \\
IRAS16355$-$2049*        & 04/23/2012 &         3$\times$1800  & ---  & ---   &      20 \\
IRAS20520$-$2329$\dagger$       & 05/02/2014    &       2$\times$1800  & ----         &  ----    & 15                    \\      
MCG$-$04.24.017*                 & 04/25/2012 &         3$\times$1800  & ---  &  ---      & 21   \\
RBS0219                 & 10/17/2015    &       3$\times$2400  & 08/18/2012     & 3$\times$1800    & 31   \\      
RHS56                   & 10/17/2015    &       4$\times$2400  & 08/18/2012     & 2$\times$1800    & 25   \\
RX J0024.7$+$0820*          & 10/26/2014 &      3$\times$2400  & ---   &        ---               & 10 \\
RX J0323.2$-$4931*          & 10/17/2015 &      4$\times$2400  & ---   &  ---              & 17 \\
RX J0902.5$-$0700*          & 04/08/2013 &      3$\times$1800  & ---   &  ---         &  8  \\
RX J2301.8$-$5508*          & 10/28/2014 &      3$\times$2400  & ---   & ---          & 23       \\
SDSS J134524.69$-$025939.8* & 04/08/2013 &      2$\times$2400   & ---   & ---          &  11   \\
SDSSJ161227.83+010159.8    & 07/13/2015  &    3$\times$1800   & 07/12/2015    & 4$\times$1800   &  5       \\      
SDSS J225452.22$+$004631.4* & 09/10/2013  &    2$\times$2400  & ---   & ---   &  10         \\
V961349$-$439*           & 04/07/2013 &         3$\times$1800  & ---   &        ---       &   19 \\
WPV8507*                 & 10/18/2015 &         3$\times$2400  & ---   &        ---       &        28  \\
Zw049.106*               & 04/29/2014 &         3$\times$2700  & ---   &        ---           &   13  \\
\hline                  
\end{tabular}
\end{table*}

\begin{table*}[!ht]
\caption{FWHM of the BC of H${\alpha}$, luminosity of that component, and black hole masses. Parameters of galaxies labeled with "*" correspond to BC$\beta$, with the black hole estimation through this line.}
\begin{center}
\begin{tabular}[h]{|l|l|l|l|l}
\hline
\hline
                         & FWHM$_{BC\alpha}$       & log L$_{BC\alpha}$  &   log(M$_{BH}$)        \\
Galaxy               & (km s$^{-1}$ )       & (erg s$^{-1}$)  &     (M$_{\odot}$)              \\  
\hline
\hline
 
    2MASXJ08173955$-$0733089  &     2697   &    41.12   &    6.7                \\
    CTSH34.06*    &    1685   &    40.23    &   5.8             \\     
    CTSM13.24   &     2715   &    41.05   &    6.7              \\         
    HE0348$-$5353   &     3065   &    42.03   &     7.3         \\
    IRAS20520$-$2329*   &     2016   &    41.82    &   6.8      \\
    RBS0219   &     2721   &    41.94   &    7.2                \\
    RHS56   &     4159   &    41.68   &    7.4          \\
    SDSSJ161227.83$+$010159.8   &     1198   &    40.66   &    5.7              \\
      
\hline
\end{tabular}
\label{tab:mass1}
\end{center}
\end{table*}

\begin{sidewaystable*}
\caption{Parameters of the [OIII]$\lambda$5007 emission profile. Column 1: Galaxy name. Columns 2 and 3: FWHM and flux of the core component of [OIII]$\lambda$5007. Columns 4 and 5: FWHM and flux of the asymmetric component of [OIII]$\lambda$5007. Column 6: Radial velocity difference between the centroid of the narrow component of H$\beta$ and the centroid of the core component of [OIII]$\lambda$5007. Column 7: Radial velocity difference between the centroid of the asymmetric component and the core component of [OIII]$\lambda$5007. Column 8: Radial velocity difference between the blue end of the blue wing and the centroid of the core component of [OIII]$\lambda$5007 (equation \ref{eq:totalasym}). Column 9: Logarithm of the Eddington ratio. All presented FWHM were corrected by instrumental broadening and are in units of km s$^{-1}$, similar to the three presented radial velocity differences. All fluxes are in units of 10$^{-15}$erg cm$^{-2}$ s$^{-1}$.}
\label{tab:oiii}
\centering            
\begin{tabular}{l c c c c c c c c }     
\hline\hline       

                  & FWHM$_{[OIII]cc}$ & Flux$_{[OIII]cc}$       & FWHM$_{[OIII]ac}$     & Flux$_{[OIII]ac}$  & $\Delta$v & $\Delta$v$_{asym}$ & total asymmetry & L$_{bol}$/L$_{Edd}$ \\
Galaxy name & (km s$^{-1}$) & (10$^{-15}$erg cm$^{-2}$ s$^{-1}$)    & (km s$^{-1}$)      & (10$^{-15}$erg cm$^{-2}$ s$^{-1}$) & (km s$^{-1}$) & (km s$^{-1}$) & (km s$^{-1}$) &  \\  \hline
 1RXSJ040443.5$-$295316   &    124  &     2.70   &    314  &      6.88  &      88  &    -173   &   -486   & 0.01  \\ 
 2MASXJ01413249$-$1528016 &    260  &     2.03   &    597  &      1.22  &     -58  &    -368   &   -965   & -0.29  \\
 2MASXJ05014863$-$2253232 &    262  &     1.35   &    761  &      12.72 &      45  &    -201   &   -963   & -0.20   \\
 2MASXJ08173955$-$0733089 &    261  &     1.18   &    1006 &      0.92  &      222 &     -123  &   -1129   & -0.93  \\
 2MASXJ21531910$-$1514111 &    415  &     1.57   &    933  &      1.01  &     -113 &    -180     &   -1113 & -0.18  \\
 2MASXJ21565663$-$1139314 &    290  &     19.04  &    795  &      5.12  &      10  &    -62   &   -857     &  0.18   \\
 CTSH34.06                &    283  &     18.75  &    571  &      7.86  &      27  &    -121   &   -692            &  0.39    \\
 CTSM02.47                &    312  &     4.13   &    369  &      1.07  &      17  &      -308   &   -677          &  -0.54   \\
 CTSM13.24              &    306  &     4.15   &    789  &      5.81  &      180 &     -153  &   -943      &  -0.35   \\ 
 FAIRALL0107              &    364  &     2.35   &    971  &      3.65  &      5   &   -35    &   -1005    &  -0.46  \\
 HE0348$-$5353            &    365  &     7.33   &    930  &      12.70  &     -3   &   -165    &   -1095  &  -0.06  \\
 HE1107$+$0129            &    270  &     4.40   &    846  &      2.47  &      9   &   -175    &   -1020   &  -0.45  \\
 IRAS04576$+$0912         &    373  &     7.98   &    1184 &      9.78  &     -220 &     -498  &   -1682   &  -0.37  \\
 IRAS16355$-$2049         &    341  &     67.50  &    940  &     101.40 &     -36  &      -224   &   -1164 &  -0.40  \\
 IRAS20520$-$2329         &    319  &     2.51   &    1010 &      3.93  &     -59  &    -267   &   -1277   &  0.36   \\
 MCG$-$04.24.017          &    203  &     17.85  &    621  &      13.81 &      7   &    -25    &   -646    & -0.60   \\
 RBS0219                  &    580  &     5.19   &    954  &      5.87  &     -1   &   -566    &   -1520   & -0.02   \\
 RHS56                    &    255  &     11.27  &    1114 &      20.09 &     23   &   -118    &   -1232   & -0.38   \\
 RXJ0024.7$+$0820         &    365  &     2.22   &    370  &      0.27  &    -43   &   -711    &   -1081   & -0.33    \\
 RXJ0323.2$-$4931         &    218  &     3.38   &    1041 &      1.69  &     176  &    -144   &   -1185   & -0.98     \\
 RXJ0902.5$-$0700         &    324  &     4.49   &    1358 &      2.75  &    -68   &   -332    &   -1689   & -0.54     \\
 RXJ2301.8$-$5508         &    476  &     6.21   &    608  &      1.24  &    -105  &      -893   &   -1501 & -0.59      \\
 SDSSJ134524.70$-$025939.8 &   336  &     4.23   &    900  &      2.92  &    -104  &    -199   &   -1099   & -0.27      \\
 SDSSJ161227.83$+$010159.8 &   236  &     0.63   &    732  &      0.93  &    -38   &    -39    &   -771    & 0.16       \\
 SDSSJ225452.22$+$004631.4 &   135  &     0.51   &    1431 &      1.96  &     288  &    -109   &   -1540   & -0.83      \\
 V961349$-$439             &   359  &     5.74   &    684 &      6.59  &    -405  &    -126   &   -810     &  -0.63     \\
 WPV85007                  &   315  &     8.72   &    1041 &      6.21  &    -186  &    -432   &   -1473   &  -0.78  \\
 Zw049.106                 &   369  &     7.53   &    905  &      3.19  &     46   &   -302    &   -1207   &  -0.38  \\
\hline                  
\end{tabular}
\end{sidewaystable*}



\begin{figure*}[!ht]
\includegraphics[trim = 65mm 14mm 0mm 10mm, clip, width=\linewidth]{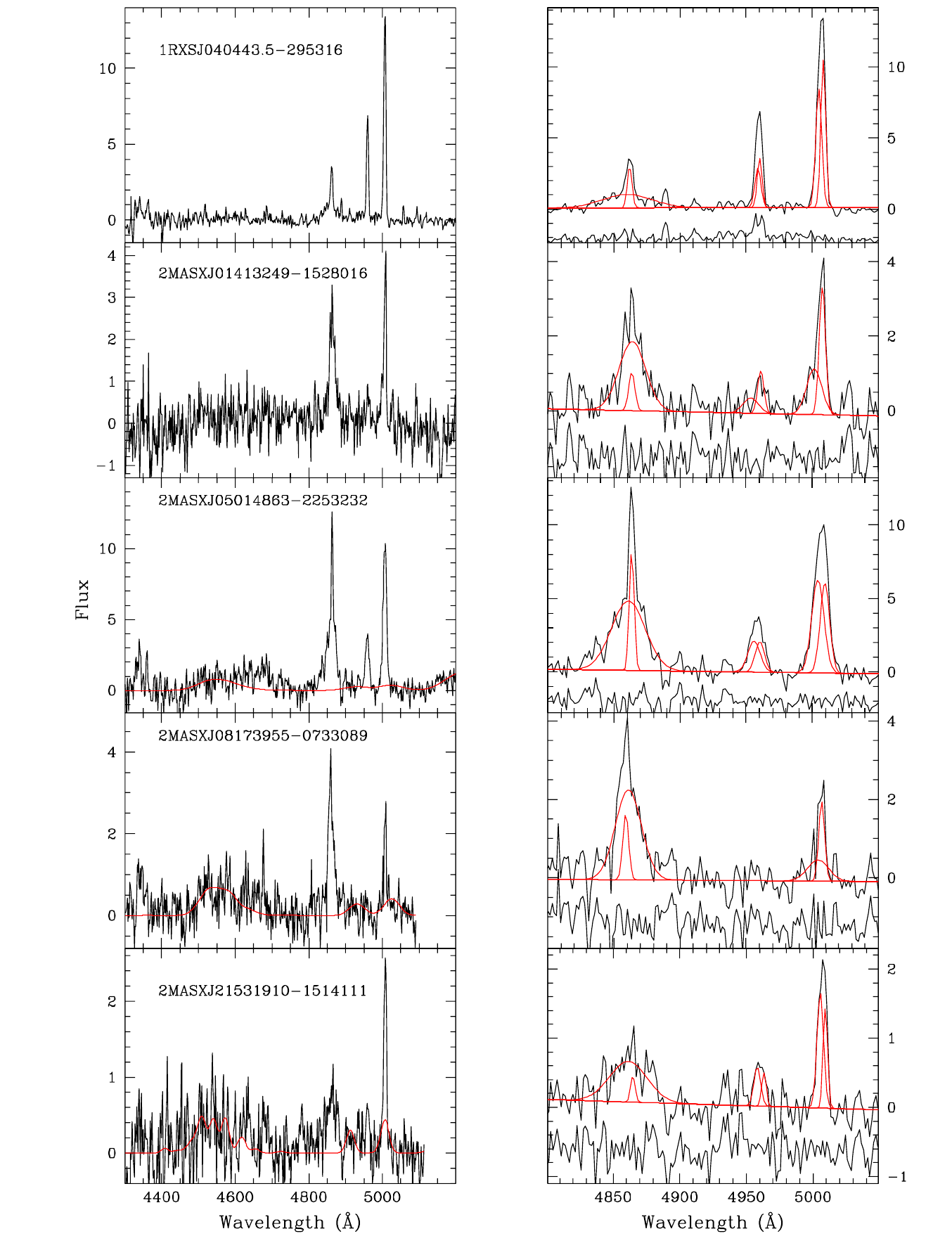}
\caption{Left: Continuum subtracted spectra of the NLS1 in the range 4300\AA{} - 5200\AA{}. Overlaid in red is the Fe multiplets fitting. Right: Gaussian decomposition of the H$\beta$ and [OIII]$\lambda\lambda$4959,5007 emission lines of the Fe subtracted spectra. The individual components are shown in red lines and the residuals are plotted at the bottom of each panel. Fluxes are in arbitrary units.}
\label{fig:spec1}
\end{figure*}

\begin{figure*}[!ht]
\ContinuedFloat
\includegraphics[trim = 65mm 14mm 0mm 10mm, clip, width=\linewidth]{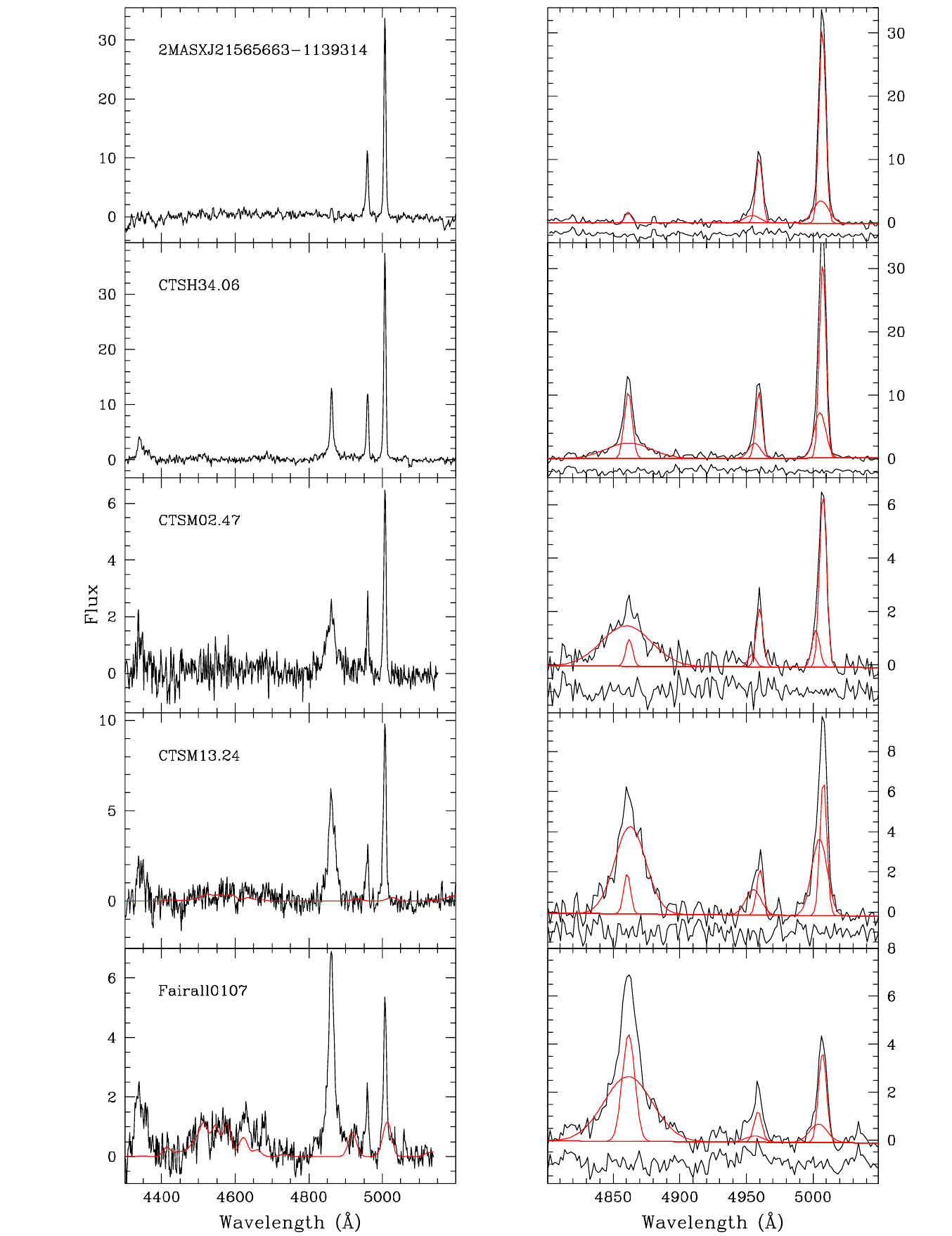}
\caption[]{Continued}
\label{fig:spec2}
\end{figure*}

\begin{figure*}[!ht]
\ContinuedFloat
\includegraphics[trim = 65mm 14mm 0mm 10mm, clip, width=\linewidth]{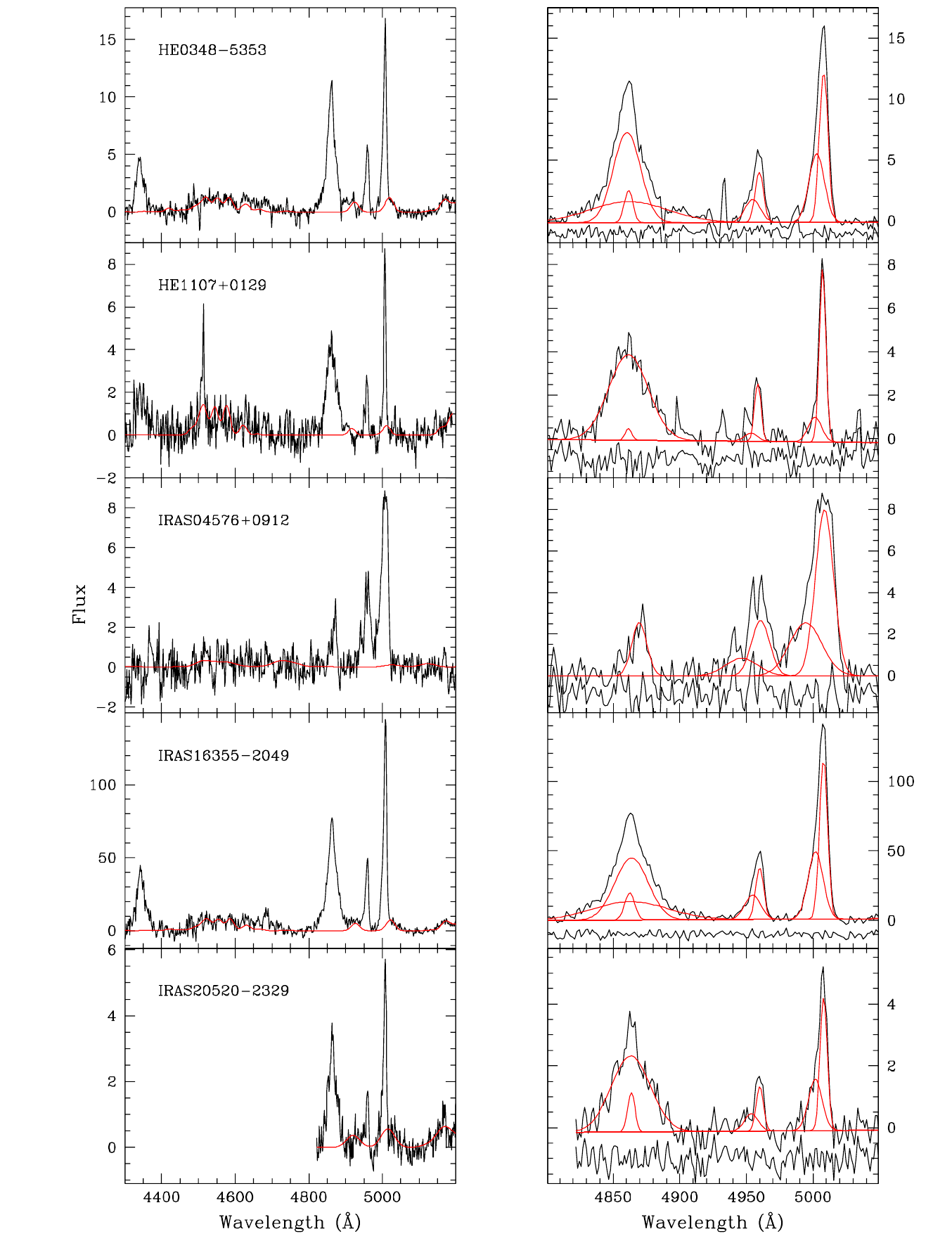}
\caption[]{Continued}
\label{fig:spec3}
\end{figure*}

\begin{figure*}[!ht]
\ContinuedFloat
\includegraphics[trim = 65mm 14mm 0mm 10mm, clip, width=\linewidth]{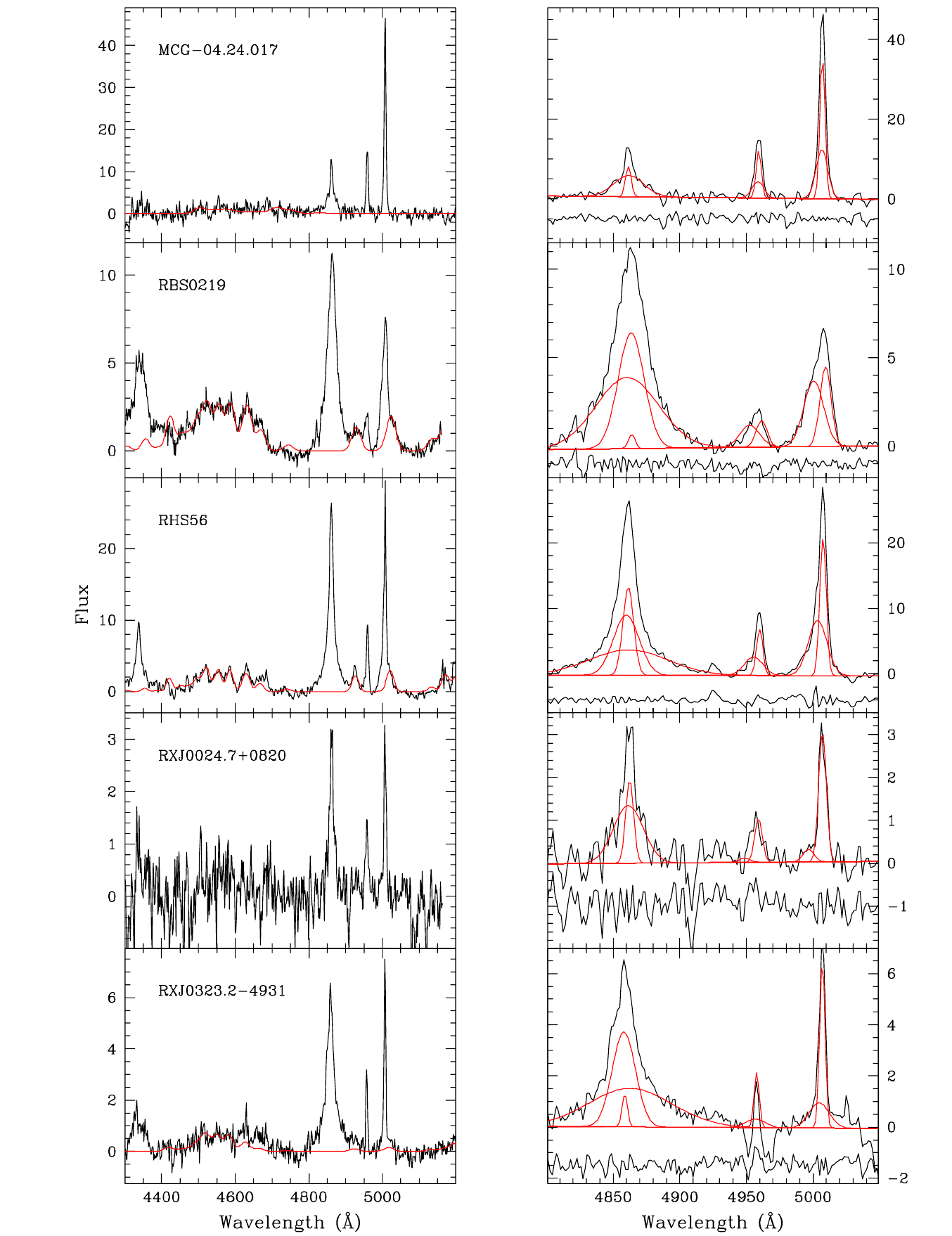}
\caption[]{Continued}
\label{fig:spec4}
\end{figure*} 

\begin{figure*}[!ht]
\ContinuedFloat
\includegraphics[trim = 65mm 14mm 0mm 10mm, clip, width=\linewidth]{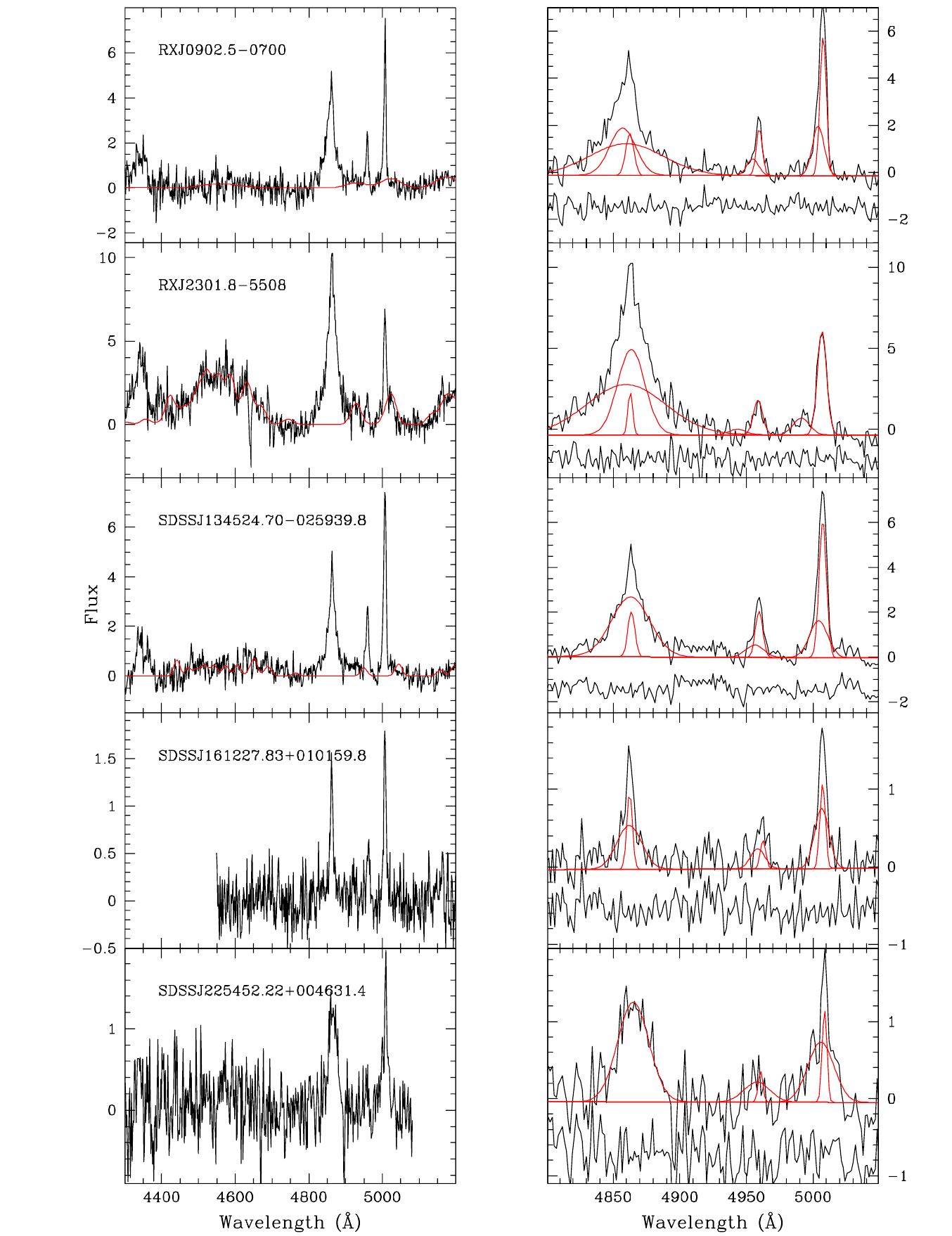}
\caption[]{Continued}
\label{fig:spec5}
\end{figure*} 

\begin{figure*}[!ht]
\ContinuedFloat
\includegraphics[trim = 65mm 77mm 0mm 10mm, clip, width=\linewidth]{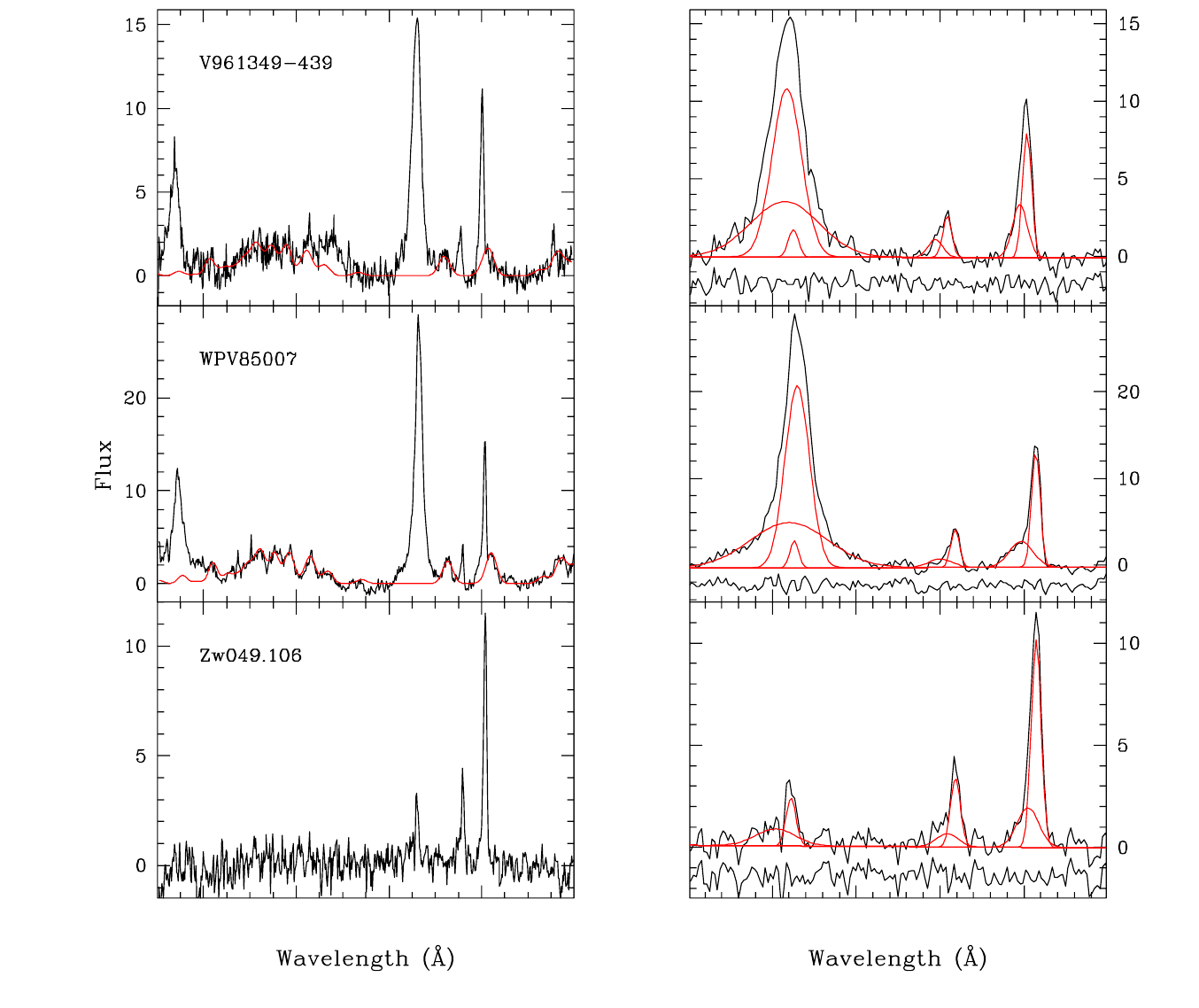}
\caption[]{Continued}
\label{fig:spec6}
\end{figure*} 

\end{document}